\documentclass[aps,nofootinbib,superscriptaddress,10pt,floatfix]{revtex4}
\normalsize
\usepackage{mathrsfs}
\usepackage{graphicx}
\usepackage{amsmath,amssymb}
\usepackage{braket}
\usepackage{subfigure}
\usepackage{subfloat}
\usepackage{float}
\usepackage[usenames]{color}
\usepackage[normalem]{ulem}
\usepackage{comment}
\usepackage{ifthen} 
\usepackage{fp,calc}
\usepackage{lastpage}
\usepackage{setspace}
\usepackage[pdftex]{epsfig}
\DeclareGraphicsExtensions{.jpg,.mps,.pdf,.png}
\usepackage{cellspace}
\usepackage{booktabs}
\usepackage{latexsym}
\usepackage{graphics}
\usepackage{fancyhdr}
\usepackage{adjustbox}
\usepackage[english]{babel}
\usepackage[svgnames]{xcolor}
\usepackage{enumitem}
\usepackage{tabularx}
\usepackage{multirow}
\usepackage[flushleft]{threeparttablex}
\usepackage{ragged2e}

\newcolumntype{C}[1]{>{\centering\arraybackslash}p{#1}}

\usepackage{titlesec}

\usepackage[colorlinks=true, allcolors=blue]{hyperref}

\usepackage{mciteplus}
 
\allowdisplaybreaks

\def\be{\begin{equation}}
\def\ee{\end{equation}}
\def\ba{\begin{eqnarray}}
\def\ea{\end{eqnarray}}

\newcommand{\HH}{\mathcal{H}}

\newcommand{\bk}{\mathbf{k}}
\newcommand{\bx}{\mathbf{x}}
\newcommand{\hJ}{\hat{J}}

\definecolor{camblue}{rgb}{0, 0, 0.55}
\definecolor{campink}{rgb}{0.85, 0.1, 0.84}

\definecolor{dgreen}{rgb}{0,0.6,0}

\newfloat{supfigure}{thp}{losf}
\floatname{supfigure}{Supplementary Figure}
\newfloat{suptable}{thp}{lost}
\floatname{suptable}{Supplementary Table}
\titleformat{\section}
  {\normalfont\fontsize{12}{14}\bfseries}{\thesection}{1em}{}
\titleformat{\subsection}
  {\normalfont\fontsize{10}{12}\bfseries}{\thesubsection}{1em}{}

\titlespacing*{\section}{0pt}{*2}{*1}
\titlespacing*{\subsection}{0pt}{*2}{*1}

\makeatletter
\renewcommand\appendix{\par
  \setcounter{section}{0}%
  \setcounter{subsection}{0}%
  \setcounter{equation}{0}%
  \gdef\@chapapp{\suppname}%
  \gdef\theequation{\thesection.\arabic{equation}}%
  \gdef\thefigure{\thesection.\@arabic\c@figure}%
  \gdef\thetable{\thesection.\@arabic\c@table}%
  \renewcommand\thesection{\suppname\ \arabic{section}}%
  \renewcommand\thesubsection{\thesection.\arabic{subsection}}}
\newcommand\suppname{Supplementary Discussion}
\makeatother

\newboolean{articletitles}
\setboolean{articletitles}{true} 
\newboolean{allauthors}
\setboolean{allauthors}{false} 

\mciteErrorOnUnknownfalse

\begin{document}

\title{Measurement of the Weyl potential evolution from the first three years of Dark Energy Survey data}
\author{Isaac Tutusaus}
\email[]{isaac.tutusaus@irap.omp.eu}
\affiliation{Institut de Recherche en Astrophysique et Plan\'etologie (IRAP), Universit\'e de Toulouse, CNRS, UPS, CNES, 14 Av.~Edouard Belin, 31400 Toulouse, France}

\author{Camille Bonvin}
\email[]{camille.bonvin@unige.ch}
\affiliation{D\'epartement de Physique Th\'eorique and Center for Astroparticle Physics, Universit\'e de Gen\`eve, Quai E. Ansermet 24, CH-1211 Geneve 4, Switzerland}

\author{Nastassia Grimm}
\email[]{nastassia.grimm@unige.ch}
\affiliation{D\'epartement de Physique Th\'eorique and Center for Astroparticle Physics, Universit\'e de Gen\`eve, Quai E. Ansermet 24, CH-1211 Geneve 4, Switzerland}

\begin{abstract}
The Weyl potential, which is the sum of the spatial and temporal distortions of the Universe's geometry, provides a direct way of testing the theory of gravity and the validity of the $\Lambda$CDM (Lambda Cold Dark Matter) model. Here we present measurement of the Weyl potential at four redshifts bins using data from the first three years of observations of the Dark Energy Survey.  We find that the measured Weyl potential is 2$\sigma$, respectively 2.8$\sigma$, below the $\Lambda$CDM predictions in the two lowest redshift bins.
We show that these low values of the Weyl potential are at the origin of the tension between Cosmic Microwave Background measurements and weak lensing measurements, regarding the parameter $\sigma_8$ which quantifies the clustering of matter. Interestingly, we find that the tension remains if no information from the Cosmic Microwave Background is used. Dark Energy Survey data on their own prefer a high value of the primordial fluctuations, together with a slow evolution of the Weyl potential. An important feature of our method is that the measurements of the Weyl potential are model-independent and can therefore be confronted with any theory of gravity, allowing efficient tests of models beyond General Relativity. 
\end{abstract}
	
\maketitle

\section{Introduction}

Since the observation of the accelerated expansion of the Universe in 1998~\cite{SupernovaSearchTeam:1998fmf,SupernovaCosmologyProject:1998vns}, extensive studies have been performed to understand if this acceleration is due to a cosmological constant, a dynamical dark energy, or a modification of the theory of gravity. In theories beyond $\Lambda$CDM, not only the background evolution of the Universe can be modified, but also the formation of structure. A powerful way of testing these theories is therefore to confront them with measurements from large-scale structure surveys. 

The landscape of theories beyond $\Lambda$CDM is vast, and testing all models one by one has become infeasible. As a consequence, two complementary approaches have been developed. 
The first one consists in testing classes of theories, e.g.\ Horndeski theories~\cite{Horndeski1974} that encompass all scalar-tensor theories with second-order equations of motion. The advantage of such an approach is that the functions constrained by the data are directly linked to fundamental ingredients in the theories. However, the degeneracies between the free functions are large, and current data are not stringent enough to constrain them all well~\cite{Noller:2018wyv}. 

The second approach is more phenomenological and consists in parameterising deviations from General Relativity (GR) directly at the level of Einstein's equations by introducing two new functions, often called $\mu$ and $\eta$~\cite{Koyama:2015vza}. The function $\mu$ encodes changes in Poisson's equation, while $\eta$ describes the difference between the time distortion and the spatial distortion of the metric, the so-called anisotropic stress. This approach has been successfully used with current data, e.g.\ combining gravitational lensing measurements from the Dark Energy Survey (DES) and galaxy clustering measurements from BOSS and eBOSS~\cite{DES:2022ygi}. Even though powerful, this approach suffers from one important complication: the constraints on $\mu$ and $\eta$ at a given redshift $z$ depend on the evolution of the functions at all redshifts above $z$. This means that either one assumes a fixed time evolution for these functions, as done in~\cite{DES:2022ygi}, and only the values of $\mu$ and $\eta$ at $z=0$ are constrained. Or one needs sophisticated techniques to constrain and reconstruct the time evolution of $\mu$ and $\eta$ from a set of chosen redshift nodes, as done for example in~\cite{Pogosian:2021mcs}. In this second case, it is necessary to have measurements over a wide range of redshift to obtain relevant constraints, and any degradation of the data in a given redshift bin will impact the constraints in the other bins. Another limitation of this framework is that the constraints on $\mu$ and $\eta$ rely on the validity of Euler's equation for dark matter. Without this additional assumption, these parameters cannot be constrained from redshift-space distortions (RSD), since they are fully degenerated with the parameters encoding changes in Euler's equation~\cite{SvevaNastassiaCamille,Bonvin:2022tii}. 

Here we propose an alternative approach, which is fully detached from any theoretical model. We ask ourselves what are the quantities that we can directly measure from the data, in a fully model-independent way. This approach has been extensively used in the past for RSD. Combining the multipoles of the RSD correlation function (or power spectrum) provides direct measurements of the galaxy peculiar velocities in the redshift bins of the surveys. The evolution of the velocity is encoded in the so-called growth rate function, $f$, which is measured in combination with $\sigma_8$ (the amplitude of density perturbations in spheres of 8 Mpc/$h$)~\cite{Song:2008qt,Blake_2011,eBOSS:2020yzd}. Such measurements are very powerful, since $f\sigma_8$ can then be compared with the prediction from GR to see if this theory is consistent or not. Moreover, it can be compared with predictions from any other theory of gravity, to put constraints on their parameters, see e.g.~\cite{eBOSS:2020yzd}. 
To the best of our knowledge, this approach, which is standard in RSD analyses, has never been used for gravitational lensing analyses with real observations. The goal of this paper is to do precisely that, using DES measurements of galaxy-galaxy lensing and galaxy clustering. More precisely, we use the formalism developed in~\cite{Tutusaus:2022cab} to measure the evolution of the Weyl potential $\Psi_W=(\Phi+\Psi)/2$ (where $\Phi$ and $\Psi$ denote respectively the spatial and temporal perturbations of the metric, see Eq.~\eqref{eq:metric} in Methods) in the four tomographic bins of the DES lenses. This provides the first direct and model-independent measurement of the evolution of the perturbed geometry of our Universe. We define a new function, $\hat{J}$, that we call the \emph{Weyl evolution}, to encode the evolution of the Weyl potential. The measurement of $\hat{J}$ from DES data is fully complementary to the growth rate measurement. Modified theories of gravity can indeed change the way structures evolve, the way the geometry evolves, or both. As for the measurement of $f\sigma_8$, the Weyl evolution $\hat{J}$ is measured redshift bin per redshift bin, without assuming any time evolution. These model-independent measurements can then be compared with the prediction in $\Lambda$CDM, or with any theory of gravity of interest.

\section{Results}

We combine two sets of data from DES: galaxy clustering and galaxy-galaxy lensing. As described in methods, subsection The lensing angular power spectra as function of the Weyl evolution, these two data sets provide constraints on cosmological parameters, nuisance parameters, as well as the galaxy bias, $\hat{b}(z_i)=b(z_i)\sigma_8(z_i)$, and the Weyl evolution, $\hat{J}(z_i)$, in each of the redshift bins of the lenses. We focus on the first four tomographic bins of the DES lens sample, with mean redshifts [0.295, 0.467, 0.626, 0.771], since some residual systematic uncertainties were identified in
the two highest redshift bins~\cite{DES:2021wwk}. We consider four different scenarios: our baseline corresponds to the use of 3$\sigma$ Gaussian priors on the cosmological parameters from Planck measurements of the Cosmic Microwave Background (CMB)~\cite{Planck:2018vyg}, while using the fiducial DES angular scale cuts. In this way, the density power spectrum at high redshift is constrained by Planck, while the late time evolution is left free. To test the sensitivity of our treatment on the non-linear modelling, we also consider a ``pessimistic'' scenario with more stringent angular scale cut, using only scales above 21\,Mpc$/h$ for the galaxy-galaxy lensing observable (see methods, subsection Measurement of the Weyl potential for more detail). Finally, we also consider two scenarios (with standard
and pessimistic scale cuts), removing the priors from Planck. In these scenarios, both the early time power spectrum and the late time evolution of the Weyl potential are constrained by DES data only. For these cases we use the wide fiducial DES priors on the cosmological parameters.

\subsection{Measurements of $\hat{J}$ using priors from Planck}

\begin{figure}
  \includegraphics[width=\linewidth]{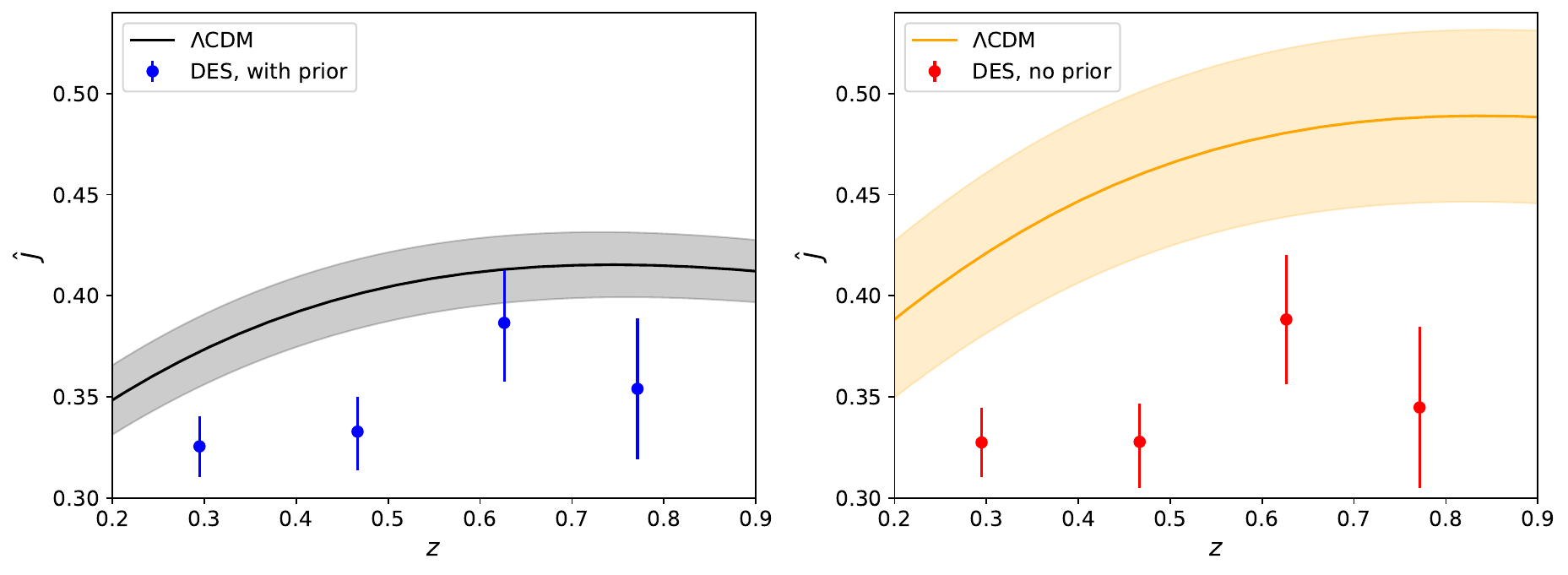}
  \caption{Measured values of $\hJ$ with 1 standard deviation error bars at the mean redshifts of the \textsc{MagLim} sample. The left
  panel shows the measurements assuming a prior from Planck on early time cosmological parameters. The right 
  panel shows the measurements without Planck priors. The black and orange solid lines show the predictions for $\hJ$ in $\Lambda$CDM. The shaded regions indicate the 1$\sigma$ uncertainty on those predictions due to uncertainties in the primordial cosmological parameters. Note that, due to the change in cosmological parameters (see Supplementary Table~1), the $\Lambda$CDM prediction of $\hJ$ changes in the case with and without prior, which is particularly noticeable at higher redshifts, where there is a $1.6\sigma$ difference between the two predictions. Source data are provided as a Source Data file.}
  \label{fig:Jhat}
\end{figure}

\begin{table}
\caption{Mean values and 1$\sigma$ error bars for $\hat{J}(z_i)$ in the four cases considered in our analysis: with and without priors from Planck measurements of the Cosmic Microwave Background (CMB), with standard and pessimistic scale cuts. We have considered a uniform prior for $\hat{J}$ between 0.1 and 0.6 for the first two bins, and between 0.15 and 0.65 for the last two bins.}  \label{Tab:Jhat}
  \begin{tabular}{lcccc}
  \toprule
     & \multicolumn{2}{c}{CMB prior} & \multicolumn{2}{c}{No CMB prior}\\\cmidrule(r){2-3}\cmidrule(l){4-5}
     & ~Standard cuts~ & ~Pessimistic cuts~  & ~Standard cuts~ & ~Pessimistic cuts~  \\
    \midrule
    $\hat{J}(z_1)$ & $0.325\pm 0.015$ & $0.329\pm 0.028$ & $0.327\pm 0.017$ & $0.335\pm 0.030$ \\
    $\hat{J}(z_2)$ & $0.333^{+0.017}_{-0.019}$ & $0.324\pm 0.035$ & $0.328^{+0.019}_{-0.023}$ & $0.326\pm 0.036$  \\
    $\hat{J}(z_3)$ & $0.387^{+0.026}_{-0.029}$ & $0.428\pm 0.061$ & $0.388\pm 0.032$ & $0.433\pm 0.063$  \\
    $\hat{J}(z_4)$ & $0.354\pm 0.035$ & $0.368\pm 0.075$ & $0.345\pm 0.040$  & $0.370\pm 0.080$ \\
    \bottomrule
  \end{tabular} 
  \end{table}

In Fig.~\ref{fig:Jhat}, left panel, we show the measured values of $\hat{J}$ with 1$\sigma$ error bars. We see that $\hat{J}$ is very well measured, with a precision of 5 to 10\% in the four DES redshift bins (see also Table~\ref{Tab:Jhat}). This demonstrates the capacity of the data to significantly constrain the Weyl evolution in a fully model-independent way, i.e.\ without assuming any theory of gravity. The measured values of $\hJ$ can then be compared with any model of interest. In Fig.~\ref{fig:Jhat}, we show the prediction for $\hat{J}$ in $\Lambda$CDM, with the cosmological parameters obtained through our cosmological analysis. The shaded region corresponds to the 1$\sigma$ uncertainty on the $\Lambda$CDM prediction. In the first and second redshift bins, the measurements of $\hat{J}$ are 2$\sigma$, respectively 2.8$\sigma$, below the $\Lambda$CDM predictions. In the highest two bins the measurements are compatible with $\Lambda$CDM (1.6$\sigma$ difference in the last bin). The joint constraints on $\hJ$ are plotted in Supplementary Fig.~1.

Even though such a tension is not very significant, it is interesting to see 
that the deviations from $\Lambda$CDM are localised to the two lowest redshift bins. This could hint either at a modification of gravity that may have more impact at low redshift, when the mechanism responsible for the acceleration of the Universe is in full play. Or it could be due to a (partially) incorrect modelling of non-linear effects, which are more important at low redshift. Alternatively, it could also be due to unknown systematic effects in the first two bins of the \textsc{MagLim} sample. 

Our measurements are in agreement with the results of~\cite{Garcia-Garcia:2021unp}, which used cosmic shear and galaxy clustering data from DES Year 1, KiDS-1000, eBOSS QSOs, DESI Legacy Survey and CMB lensing to reconstruct the redshift evolution of matter density perturbations. Their methodology differs from ours, first because they use GR to link the Weyl evolution to density perturbations, allowing them to combine cosmic shear, CMB lensing and galaxy clustering to measure a single growth function, and second because they reconstruct the growth function using spline interpolation between chosen redshift nodes. As can be seen from Fig.~8 of~\cite{Garcia-Garcia:2021unp}, they find a growth function which is lower than predicted by $\Lambda$CDM between $z=0.2$ and $z=0.5$, in perfect agreement with our measurements of a low Weyl evolution in the first two redshift bins. Both methods are highly complementary: using GR and spline interpolation allows them to combine different probes, enhancing the precision of their measurement. Our method is more model-independent since it does not rely on GR, which allows us to compare our measurements of $\hJ$ with the predictions of various modified gravity models, see subsection Modified gravity below. Moreover, by measuring $\hJ$ at the redshifts of the lenses, our method can also directly identify if the tension comes from one particular redshift bin (which could indicate a problem with the lens sample in that bin), or from various redshift bins: the two lowest ones in our case. In Supplementary Discussion 2, we also compare our results with the binned $\sigma_8$ analysis of DES and we show that the two approaches are complementary but different: the binned $\sigma_8$ analysis is a consistency test of the $\Lambda$CDM model, whereas our approach provides a model-independent way of measuring the Weyl evolution. For completeness, in Supplementary Discussion 3, we show and discuss the results for the \textsc{redMaGiC} lens sample of DES~\cite{DES:2015pcw}.

Since $\hat{J}$ has been measured in a fully model-independent way, we can now use these measurements to infer $\sigma_8(z=0)$ assuming $\Lambda$CDM. Combining the four measurements of $\hat{J}$, we can fit for $\sigma_8(z=0)$ at each point of the chain using Eq.~\eqref{eq:Jhat_sigma8} in methods, and obtain its posterior distribution which we find to be close to Gaussian (see Supplementary Fig.~2). The mean and 1$\sigma$ errors are $\sigma^{\rm fit}_8(z=0)=0.743\pm0.039$. In contrast, using the values of $\Omega_{\rm m},\Omega_{\rm b}, n_{\rm s}$ and $A_{\rm s}$, we obtain $\sigma^{\rm cosmo}_8(z=0)=0.849\pm 0.030$, which is 2.2$\sigma$ higher than the value obtained from $\hat{J}$. With our method, we recover therefore the well-known result from the Year 3 DES data that high redshifts prefer a larger clustering amplitude than low redshifts. Moreover, we are able to pin-point this tension to the behaviour of the Weyl potential in the two lowest redshift bins of DES. 
We obtain similar results if we use $\hJ$ to infer $S_8=\sigma_8 \sqrt{\Omega_{\rm m}/0.3}$: the value obtained from $\hJ$ is 2.1$\sigma$ below that obtained from cosmological parameters. 

Our results are in agreement with previous analyses, which combined CMB lensing (either from ACT or from Planck) with clustering from DES~\cite{ACT:2023ipp}, BOSS~\cite{Chen:2022jzq}, DESI~\cite{White:2021yvw} and high-redshift quasars~\cite{Alonso:2023guh,Piccirilli:2024xgo}. These analyses all point toward a clustering amplitude which is consistent with $\Lambda$CDM above redshift 0.5~\cite{Alonso:2023guh,Piccirilli:2024xgo,ACT:2023kun}, but is too small at low redshift~\cite{White:2021yvw,Chen:2022jzq,ACT:2023ipp,Garcia-Garcia:2021unp}. This shows that the tension is not limited to cosmic shear measured in galaxy surveys, but also appears when using CMB lensing. Combined with the fact that some of those analyses used BOSS and DESI galaxy samples instead of the DES lens sample, we can conclude that the tension is not linked to a particular data set. Interestingly, these results are also in agreement with other probes, e.g.\ the measurements in~\cite{Esposito:2022plo}, which find that high-redshift Lyman-$\alpha$ data prefer a large value of $\sigma_8$ compared to the number counts of clusters at low redshift pointing towards a lower $\sigma_8$.

\begin{table}
\caption{Mean values and 1$\sigma$ error bars for $\sigma_8(z=0)$, in the cases with and without CMB priors, using the standard scale cuts. We compare the value fitted from $\hJ$ with the value obtained from the cosmological parameters.} \label{Tab:sigma8}
  \begin{tabular}{lcccc}
  \toprule
     & \multicolumn{2}{c}{CMB prior} & \multicolumn{2}{c}{No CMB prior}  \\\cmidrule(r){2-3}\cmidrule(l){4-5}
     & From parameters & From $\hat{J}$  & From parameters & From $\hat{J}$  \\
    \midrule
    $\sigma_8(z=0)$ & ~$0.849\pm0.030$~ & ~$0.743\pm 0.039$~ &~$1.028\pm 0.097$~ & ~$0.776^{+0.067}_{-0.079}$~ \\
    \bottomrule
\end{tabular} 
\end{table}

\subsection{Measurements of $\hat{J}$ without Planck prior}

We then redo the measurements of $\hat{J}$, removing the priors from Planck. From Fig.~\ref{fig:Jhat} (right panel), we see that $\hat{J}$ can still be very well measured in this case: the errors increase by $11-20\%$ only (see Table~\ref{Tab:Jhat}). On the other hand, the errors on the cosmological parameters are 3 to 6 times larger when no priors are used ($A_{\rm s}$ and $h$ being the most degraded). 
This degradation is not surprising, since in our framework all the information coming from the growth of structure and from the Weyl potential is encoded in the parameters $\hat{b}_i$ and $\hat{J}$ and does not contribute to the constraints on the cosmological parameters (see methods, subsection The lensing angular power spectra as function of the Weyl evolution for more detail). The larger uncertainties in cosmological parameters lead to wider errors on the prediction for $\hJ$ in $\Lambda$CDM, as is clearly visible from Fig.~\ref{fig:Jhat}. 

We see that removing the priors actually slightly increases the tension between the measured $\hJ$ and the $\Lambda$CDM predictions, which is now $[2.1\sigma, 2.9\sigma, 1.7\sigma, 2.4\sigma$] in the four bins. This is due to the fact that the $\Lambda$CDM predictions for $\hJ$ increase when the prior is removed, due to an increase of the primordial amplitude $A_{\rm s}$ from $\left(\,2.13\pm 0.10\,\right)\cdot 10^{-9}$ to $\left(\,2.98^{+0.57}_{-0.69}\,\right)\cdot 10^{-9}$. In this case, $A_{\rm s}$ is mainly constrained by the (subdominant) RSD contribution to galaxy clustering that we modelled in GR (see methods, subsection Measurement of the Weyl potential and Supplementary Discussion~1 for more detail).  We note that the value of $A_{\rm s}$ is still compatible with Planck value at 1.2$\sigma$, but its slight increase counterbalances the increase of uncertainty due to the absence of priors, leading to a persisting tension of $\hJ$ with $\Lambda$CDM predictions.  From this analysis, we see that DES data on their own prefer a high amplitude of perturbations $A_{\rm s}$, together with a growth of the Weyl potential slower than in $\Lambda$CDM at low redshifts.

The tension in $\hJ$ leads to a 2.1$\sigma$ tension between $\sigma_8$ extracted from the cosmological parameters: $\sigma^{\rm  cosmo}_8(z=0)=1.028\pm 0.097$, and $\sigma_8$ extracted from $\hJ$ assuming $\Lambda$CDM: $\sigma^{\rm fit}_8(z=0)=0.776^{+0.067}_{-0.079}$. The large value of $\sigma^{\rm cosmo}_8$ is directly related to the large value of $A_{\rm s}$. 
Our analysis shows therefore that the $\sigma_8$ tension is internally present in DES data. The clustering amplitude extracted from the evolution of the Weyl potential is in tension with the clustering amplitude inferred from the background evolution and the subdominant RSD signal in the distribution of lenses. Our method therefore indicates that the $\sigma_8$ tension is not only a tension between high redshift and low redshift, but that it is also a tension between the perturbations of the geometry at low redshift (probed by weak lensing), and the other dynamical fields: expansion rate and density fluctuations (encoded in the RSD contribution). This is in agreement with measurements of the growth rate function $f$ from spectroscopic RSD surveys, which show no deviations from $\Lambda$CDM at low redshift (see e.g.\ Fig.\ 1 of~\cite{Grimm:2024fui}). What is particularly interesting in our method is that it allows the separation of information from the Weyl potential and from background and density perturbations within DES data, consequently showing that the same trend is present in this single data set. Translating our constraints on $\sigma_8(z=0)$ to constraints on $S_8$ we find that the tension is slightly larger, of 2.6$\sigma$. We summarise the values of $\sigma_8(z=0)$ obtained in various cases in Table~\ref{Tab:sigma8}.

\begin{figure}
  \includegraphics[width=\linewidth]{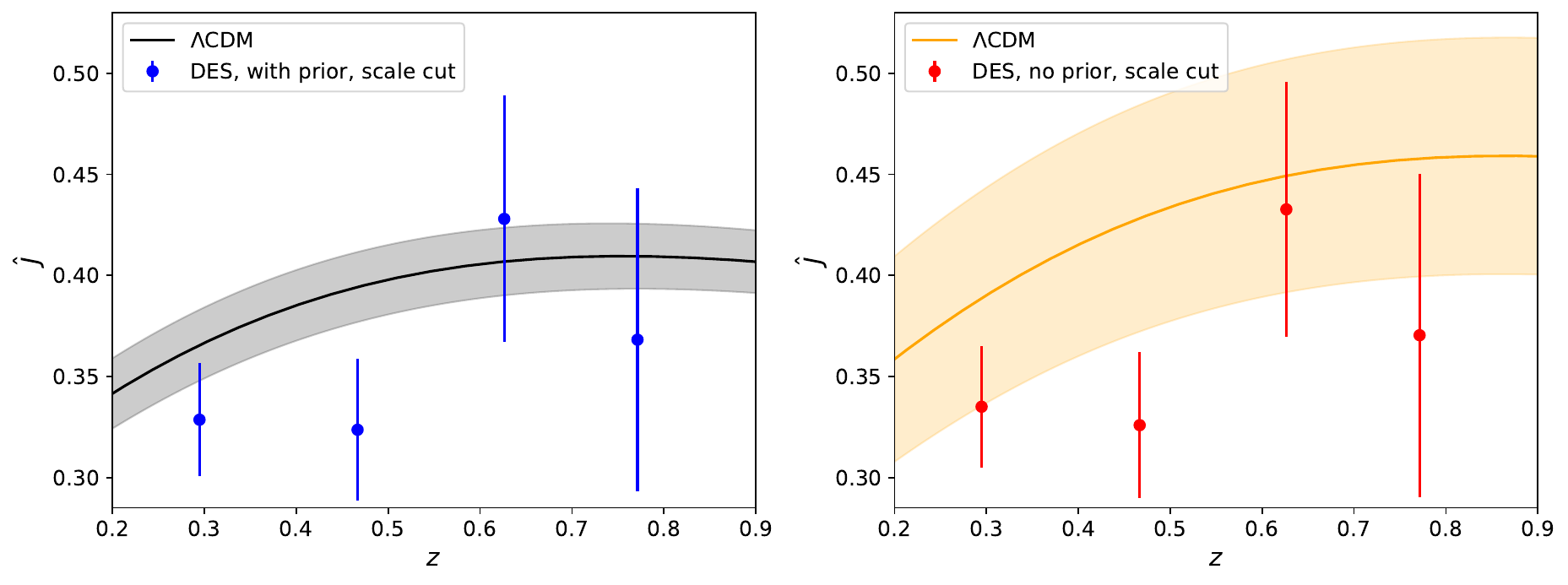}
  \caption{Measured values of $\hJ$ with 1 standard deviation error bars at the mean redshifts of the \textsc{MagLim} sample, using more stringent scale cuts. The left
  panel shows the measurements assuming a prior from Planck on early time cosmological parameters. The right
  panel shows the measurements without Planck priors. The black and orange solid lines show the predictions for $\hJ$ in $\Lambda$CDM. The shaded regions indicate the 1$\sigma$ uncertainty on those predictions due to uncertainties in the primordial cosmological parameters. Source data are provided as a Source Data file.}
  \label{fig:Jhat_cut}
\end{figure}

\subsection{Measurements of $\hat{J}$ with more stringent non-linear scale cuts}

As discussed in methods, subsection Measurement of the Weyl potential, we have implemented more stringent scale cuts to further remove the impact of non-linear scales (see Fig.~\ref{fig:gammat} in methods). The results for $\hJ$ are shown in Fig.~\ref{fig:Jhat_cut} for the analysis with Planck priors (left panel) and without (right panel). We see that the measurements of $\hJ$ are significantly degraded: the errors increase by roughly a factor 2. This is not surprising since removing scales reduces the available amount of information needed to measure $\hJ$. The errors on the $\Lambda$CDM prediction also increase due to the scale cut, in the case without prior, but only by 14\%. Interestingly, the mean values of $\hJ$ are only mildly affected by removing scales and are still below the $\Lambda$CDM predictions. However, the increase in the error bars significantly reduces the tension with $\Lambda$CDM predictions: only the second bin is still 1.8$\sigma$ away from the $\Lambda$CDM prediction in the case with prior (1.5$\sigma$ without prior). A comparison of the measurements of $\hat{J}$ with standard and pessimistic scale cuts is provided in Table~\ref{Tab:Jhat}.

\subsection{Modified gravity} \label{sec:modgrav}

\begin{figure}
  \includegraphics[width=9cm]{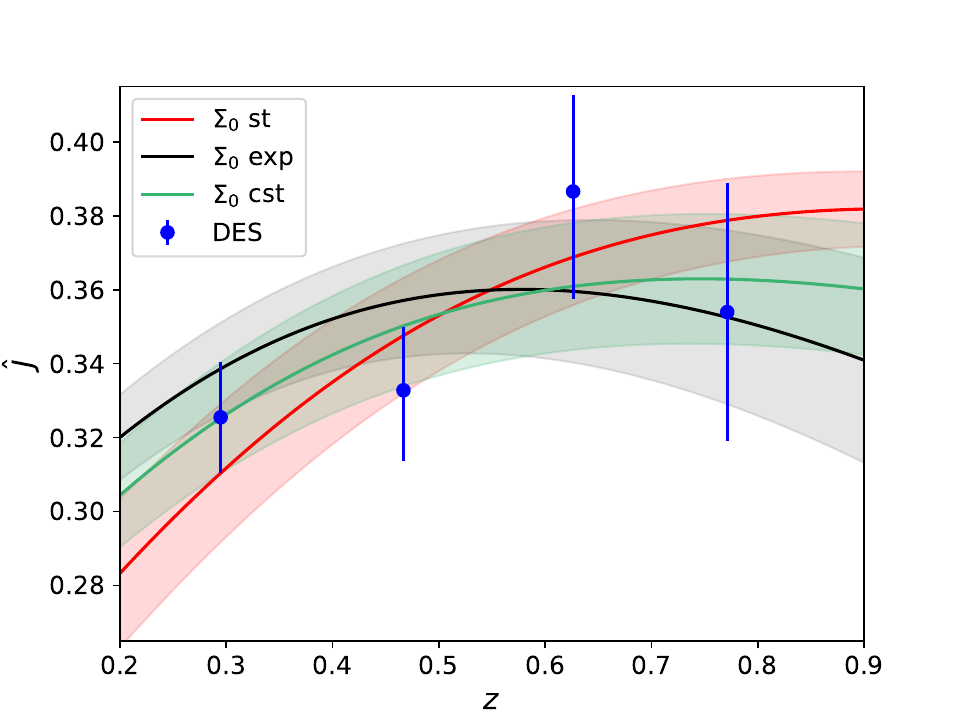}
  \caption{Predictions for $\hJ$ for modified gravity models with $\Sigma\neq 1$ and $\mu=1$. We show the mean value together with the 1 standard deviation error bars for the three choices of time evolution (standard, constant, and exponential) described in Methods. The measurements correspond to our baseline, using Planck priors and with standard scale cuts. Source data are provided as a Source Data file.}
  \label{fig:Jhat_MG}
\end{figure}

\begin{figure}
  \includegraphics[width=10.4cm]{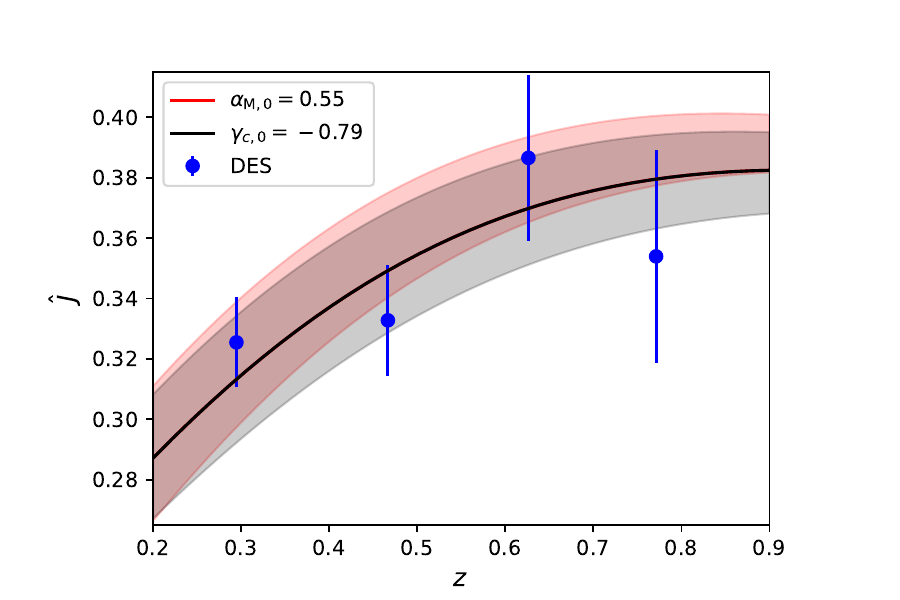}
  \caption{Predictions for $\hJ$ in Horndeski models. We show the mean value and 1 standard deviation error bars for a model with $\alpha_{\rm M}=2\alpha_{\rm B}$ (red) and a model with a non-minimal coupling, $\gamma_c\neq 0$, of CDM to the metric (black). In both cases, all other modified gravity parameters are assumed to maintain their GR values. The measurements correspond to our baseline, using Planck priors and with standard scale cuts. Source data are provided as a Source Data file.} 
  \label{fig:Jhat_Horndeski}
\end{figure}

We now compare our measurements of $\hJ$ in the baseline scenario with predictions in modified theories of gravity.
We explore two cases: A) the phenomenological $\mu-\Sigma$ extensions of gravity; and B) Horndeski theories, concentrating on a few specific cases. The relation between $\hJ$ and the parameters of these two types of modified gravity models are given in methods, subsection Modified theories of gravity. In both cases, we use the measured $\hJ$ to constrain the parameters of the theory, and then use those to compute the best fit $\hJ(z)$.

In case A), we fix $\mu=1$, since as explained in methods $\hJ$ is mainly affected by $\Sigma$. We explore three types of evolution for the function $\Sigma(z)$: a standard evolution, where $\Sigma(z)$ decays proportionally to dark energy; no evolution; and an exponential evolution (see Eq.~\eqref{eq:sigmaevol} in methods). The posteriors for $\Sigma_0=\Sigma(z=0)$ are very close to Gaussian for the three cases, with mean and 1$\sigma$ errors given by: $\Sigma_0^{\rm st}=-0.24\pm 0.10$ (standard evolution), $\Sigma_0^{\rm cst}=-0.13 \pm 0.06$ (no evolution) and $\Sigma_0^{\rm exp}=-0.027\pm 0.013$ (exponential evolution).
In Fig.~\ref{fig:Jhat_MG} we show the predicted $\hJ(z)$ (mean and $1\sigma$ error bands) for the three models, compared with the measured $\hJ$ in the baseline case. In the three models, a negative value of $\Sigma_0$ decreases $\hJ$ with respect to $\Lambda$CDM, leading to a better agreement with the data (at the cost of one extra parameter though). The reduced chi-squared of $\Lambda$CDM is of 2.1 ($p$-value 0.078), while the reduced chi-squared of the three models (with one extra free parameter $\Sigma_0$ each) is very similar: 1.1 for the standard evolution and the constant case ($p$-value 0.33), and 0.8 for the exponential evolution ($p$-value 0.48). This shows that the data are not able to discriminate between the different time evolutions, as is clear from Fig.~\ref{fig:Jhat_MG}.  The coming generation of surveys has clearly the potential to help, by adding measurements at lower and higher redshifts and reducing the error bars. Here we also see the strong advantage of our method: by first measuring $\hJ$ without assuming any model and then comparing with the predictions in modified gravity, we can easily understand and test the impact of different assumptions about the time evolution.

In case B) we consider Horndeski models while additionally allowing for a non-minimal coupling of dark matter, which constitutes a breaking of the weak equivalence principle. We focus on two specific cases: first, we consider the case where $\alpha_{\rm B}=\alpha_{\rm M}/2$ can differ from zero, and dark matter is minimally coupled to the scalar field, $\gamma_c=0$ (see e.g.\ \cite{Castello:2023zjr} for more detail on these parameters). This special relation between $\alpha_{\rm M}$ and $\alpha_{\rm B}$ is recovered e.g.~in Brans-Dicke theories~\cite{Brans:1961sx, Sawicki:2012re} and some $f(R)$ models~\cite{Carroll:2003wy, Song:2006ej, Carroll:2006jn, Vollick:2003aw}. Second, we investigate the case with a non-minimal coupling of dark matter, $\gamma_c\neq 0$, while all other parameters are kept to their GR value. For the two specific models, we obtain $\alpha_{\rm M,0}=2\alpha_{\rm B,0}=0.55^{+0.22}_{-0.20}$ and $\gamma_{c,0}=-0.79^{+0.30}_{-0.37}$. The results for $\hJ$ are shown in Fig.~\ref{fig:Jhat_Horndeski}. We see that both $\alpha_{\rm M}=2\alpha_{\rm B}>0$ as well as $\gamma_c<0$ can decrease the amplitude of the Weyl potential at late time, leading to predicted values of $\hat{J}(z)$ in line with the measured data. This shows that while $\hJ$ provides an excellent test of the $\Lambda$CDM model, it cannot, on its own, distinguish between modifications of gravity and a breaking of the weak equivalence principle. A combination with other observables, in particular gravitational redshift, will allow to break degeneracies between different physical effects~\cite{Castello:2023zjr}.

\section{Discussion}
\label{sec:conclusion}

In this paper, we have performed the first measurement of the Weyl potential from galaxy-galaxy lensing. Using data from DES, we were able to measure the function $\hJ(z)$, which encodes the evolution of the Weyl potential, with a precision of 5 to 10\% in the four bins of the \textsc{MagLim} lens sample. We found that the measured values in the two lowest bins are in mild tension with the $\Lambda$CDM predictions. 

One important feature of our method is that it allows us to separate the information coming from the matter density fluctuations (which we do not use to constrain gravity, since it is fully degenerated with the galaxy bias) from the information coming from the Weyl potential. The latter purely encodes the growth of the spatial and temporal distortions of the geometry, and is therefore independent of the galaxy bias. Moreover, our method does not rely on any assumption about the redshift evolution of $\hJ$: the function is directly measured in each bin of the survey. 

Our measurements allow us to pin-point the $\sigma_8$ tension to the fact that the Weyl potential is smaller than predicted in $\Lambda$CDM in the two lowest bins of the DES data. Surprisingly, we found that the tension remains even when no priors from the CMB are used. DES data on their own prefer a Universe with a high initial value of the primordial potential (higher than measured by Planck) followed by a slower growth at late time.  Further investigation to understand the origin of this behaviour is needed. In particular, it would be interesting to redo the analysis changing the settings to the ones used in the new combined analysis of DES and KiDS~\cite{Kilo-DegreeSurvey:2023gfr} (which found a decrease of the $\sigma_8$ tension) to see how this impacts the measurements of $\hJ$. Moreover, having measurements of $\hJ$ between the second and third bin of DES, where the deviation suddenly appears could help understand the origin of the tension. Surveys like Euclid and LSST will be of crucial importance for this. 

More generally, our method allows us to compare the measurements of $\hJ$ with predictions in any theory of gravity that recovers GR at high redshift. Since the measurements are model-independent, the method allows for various tests of the theory of gravity in an efficient way. One can for example easily change the time evolution of the parameters of the theory, without having to reanalyse the data for each case. Using the $\mu-\Sigma$ parameterisation of modified gravity, we have shown that current data are not constraining enough to differentiate between different time evolutions. In the future, we will study if correlations of galaxy clustering with CMB lensing may help. Moreover, future surveys like Euclid and LSST are expected to provide sufficiently tight constraints (at the level of $0.1-1\%$ for $z\in [0.25,2.1]$ for LSST~\cite{Tutusaus:2022cab}) over a wider redshift range, to allow for much more stringent tests. 

The evolution of the Weyl potential can be combined with measurements of the growth rate of structure, to test the relation between these two fundamental quantities in our Universe. In~\cite{Grimm:2024fui}, we have combined these quantities to obtain precise measurements of the $E_G$ statistic, providing a robust test of GR. Therefore, while $\hat{J}$ reveals an interesting tension with $\Lambda$CDM on its own, it is particularly powerful in combination with further model-independent observables.

\section{Methods}

The Weyl evolution $\hJ$ is measured from galaxy-galaxy lensing, combined with galaxy clustering. The formalism to relate gravitational lensing observables to $\hat{J}$ has been derived in~\cite{Tutusaus:2022cab}. Here we summarise the main results. 

\subsection{The lensing angular power spectra as function of the Weyl evolution}
\label{sec:theory}

We work with the linearly perturbed flat Friedmann-Lema\^itre-Robertson-Walker (FLRW) metric in the conformal Newtonian gauge, with the line element given by
\begin{align}
\label{eq:metric}
 \mathrm ds^2=a^2(\tau)\Big[-(1+2\Psi)\mathrm d\tau^2+(1-2\Phi)\mathrm d\bx^2 \Big]\, , 
\end{align}
where $\tau$ denotes conformal time, $a$ is the scale factor, and $\Psi$ and $\Phi$ are the two metric potentials. 
Gravitational lensing is directly sensitive to the Weyl potential $\Psi_W\equiv(\Phi+\Psi)/2$. We define the Weyl transfer function $T_{\Psi_W}$, which relates the value of $\Psi_W(z,\bk)$ in Fourier space and at redshift $z$ to the primordial potential at the end of inflation $\Psi_{\rm in}(\bk)$. In GR, one can easily show that for wavelengths that are inside the horizon, the Weyl transfer function simply evolves proportionally to $D_1(z)\Omega_{\rm m}(z)$. Here $D_1(z)$ is the growth function (governing the evolution of matter density perturbations) and $\Omega_{\rm m}(z)$ is the matter density parameter at redshift $z$. Hence, in GR, measuring the evolution of density perturbations or measuring the evolution of the Weyl potential provides exactly the same information. In modified theories of gravity the situation is different: Einstein's equations are modified, generically leading to a different evolution of the Weyl potential. To capture this, we introduce a new function, $J$, encoding this evolution. More precisely, we write the transfer function of the Weyl potential as (see~\cite{Tutusaus:2022cab} for more detail)
\begin{align}\label{eq:evolJ}
    T_{\Psi_W}(k,z)=&\frac{\mathcal{H}^2(z)J(k,z)}{\mathcal{H}^2(z_*)D_1(z_*)}\frac{\sqrt{B(k,z)}}{\sqrt{B(k,z_*)}}T_{\Psi_W}(k,z_*)\,,
\end{align}
where $z_*$ is a redshift well in the matter era (before the acceleration of the Universe started), $\mathcal{H}$ is the Hubble parameter in conformal time, and $B(k,z)$ is a boost factor, encoding the non-linear evolution of matter density perturbations at small scales. We assume that at $z_*$ a cold dark matter (CDM) Universe, governed by GR, is recovered. This assumption automatically excludes early dark energy models, where deviations from a CDM Universe take place already at high redshift. The motivation behind this assumption is that, at early times, CMB measurements are highly compatible with a CDM Universe~\cite{Planck:2018vyg}. Theories of modified gravity therefore usually aim at recovering GR at high redshift, while leading to a phase of accelerated expansion at low redshift. In our framework, any deviation in the Weyl potential evolution during the phase of accelerated expansion is encoded in the free function $J$. By measuring this free function as a function of redshift, we are able to reconstruct the Weyl evolution and assess its compatibility with GR predictions, or with the predictions of any theory of gravity we may want to test. Note that the function $J$ has the same information content as the function $L$ defined in~\cite{Amendola:2012ky,Amendola:2013qna}, 
which identified for the first time the quantities that can be measured in a model-independent way from large-scale structure surveys.

In full generality, $J$ is a function of redshift and wavenumber $k$. In GR, the evolution is almost scale-independent, except from a small scale-dependence due to massive neutrinos. In our analysis we fix the sum of neutrino masses to 0.06 eV, leading to a scale-dependence in the growth of structure (and therefore in $J$) which is negligible~\cite{Euclid:2019clj}. In models of gravity beyond GR, $J$ can in principle depend on scales. However, in many models of modified gravity, it turns out that the evolution is scale-independent for sub-horizon modes, where the quasi-static approximation is valid~\cite{Gleyzes:2015rua,Raveri:2021dbu}. In our analysis, we therefore drop the scale-dependence in $J$ (as is also usually done for measurements of the growth rate in RSD analyses~\cite{Song:2008qt,Blake_2011,eBOSS:2020yzd}). This simplifies the analysis, but it is not a fundamental limitation of the method, which can be extended to a scale-dependent $J$. 

As observable we consider the galaxy-galaxy lensing angular power spectrum, i.e.\ the cross-correlation of cosmic shear of background galaxies with clustering of foreground lenses. As shown in~\cite{Tutusaus:2022cab}, the galaxy-galaxy lensing angular power spectrum can be written in the following way (using the Limber approximation):
\begin{align}
    C_{\ell}^{\Delta\kappa}(z_i,z_j)=&\frac{3}{2}\int\text{d}z\,n_i(z)\mathcal{H}^2(z)\hat{b}_i(z)\hat{J}(z)
    B\left(k_{\ell},\chi\right)\frac{P_{\delta\delta}^{\rm lin}\left(k_{\ell},z_*\right)}{\sigma_8^2(z_*)}
    \int\text{d}z'n_j(z')\frac{\chi'(z')-\chi(z)}{\chi(z)\chi'(z')}\,, \label{eq:DeltakappaJ}
\end{align}
where $n_i$ and $n_j$ denote the galaxy distribution function of the lenses and sources, respectively, and $k_\ell\equiv (\ell+1/2)/\chi$, with $\chi$ the comoving distance. $P_{\delta\delta}^{\rm lin}$ is the linear matter power spectrum, evaluated here at redshift $z_*$. We see that the galaxy-galaxy lensing angular power spectrum depends on: 1) the density fluctuations at $z_*$ where GR is recovered; 2) the evolution of background quantities, $\HH(z)$ and $\chi(z)$; and 3) the functions $\hat{J}$ and $\hat{b}_i$ defined as
\begin{align}
    \hat{J}(z)&\equiv\frac{J(z)\sigma_8(z)}{D_1(z)}=\frac{J(z)\sigma_8(z_*)}{D_1(z_*)}\,, \label{eq:Jhat}\\
    \hat{b}_i(z)&\equiv b_i(z)\sigma_8(z)\, ,\label{eq:bhat}
\end{align}
where $b_i$ is the linear galaxy bias in the tomographic redshift bin $i$. The second equality in Eq.~\eqref{eq:Jhat} follows from the fact that $\sigma_8(z)$ is directly proportional to $D_1(z)$. As a consequence $\hJ$ and $J$ contain the same information about modified gravity, since they are related by the ratio $\sigma_8(z_*)/D_1(z_*)$ which does not depend on the modifications of gravity (since we assume that at $z_*$ GR is recovered). In the following we will assume the background evolution to be as in $\Lambda$CDM. This is an assumption that is often done in large-scale structure analyses, since all current constraints set the background evolution to be close to that of $\Lambda$CDM. This assumption can be relaxed, but it requires adding other data sets in the analysis (e.g.\ supernovae or BAO data) that would constrain the evolution of these functions. This is, however, beyond the scope of this analysis. 

Since $\hat{b}_i(z)$ and $\hJ(z)$ vary slowly with redshift, we can take them out of the integral in Eq.~\eqref{eq:DeltakappaJ} and evaluate them at the mean redshift of the bin $z_i$. We see therefore, that by measuring the galaxy-galaxy lensing angular power spectrum for lenses at redshift $z_i$, we can directly measure the functions $\hat{J}$ and $\hat{b}_i$ at that redshift, in a fully model-independent way, i.e.\ without assuming any theory of gravity, nor specifying a redshift evolution for $\hJ$. The fact that galaxy-galaxy lensing depends on $\hJ$ at the redshift of the lenses follows from the fact that, in the Limber approximation, the signal is fully due to correlations between the Weyl potential and the galaxy density at the position of the lenses. All the other correlations along the photon's trajectory are negligible. The situation for the shear angular power spectrum is different, since it depends on the integral of $\hJ$ from the sources to the observer. For this reason we do not consider it in our analysis. Note however that this could be done by introducing a function $\hJ$ which is piece-wise constant in some chosen redshift bins, with smooth interpolation between them.

Eq.~\eqref{eq:DeltakappaJ} shows that there is a degeneracy between $\hJ$ and $\hat{b}_i$, but this degeneracy can be easily broken by adding the clustering angular power spectrum which only depends on $\hat{b}_i$. Following~\cite{DES:2021bpo}, 
we use the Limber approximation to compute the galaxy clustering spectrum at large $\ell$ 
\begin{align}\label{eq:DeltaDeltaJ}
    C_{\ell}^{\Delta\Delta}(z_i,z_j)=&\int\text{d}z\,n_i(z)n_j(z)\frac{\mathcal{H}(z)(1+z)}{\chi^2(z)}\hat{b}_i(z)\hat{b}_j(z)
     B\left(k_{\ell},\chi\right)\frac{P_{\delta\delta}^{\rm lin}\left(k_{\ell},z_*\right)}{\sigma_8^2(z_*)}\,, \quad \mbox{for $\ell\geq 200$}\,.
\end{align}
For low $\ell$, however, the Limber approximation is not accurate enough and we need to go beyond it. Following~\cite{Fang:2019xat}, we split the non-linear matter power spectrum into its linear contribution and the difference between the non-linear and linear contributions: $P_{\delta\delta}^{\rm nl}=P_{\delta\delta}^{\rm lin}+(P_{\delta\delta}^{\rm nl}-P_{\delta\delta}^{\rm lin})$. For the purely non-linear part, $P_{\delta\delta}^{\rm nl}-P_{\delta\delta}^{\rm lin}$, the Limber approximation can be used, since non-linearities will only contribute at small scales (large $k$) where the Limber approximation is valid. The linear part on the other hand is computed exactly for $\ell<200$. In our formalism we can write it as (this corresponds to the second term of Eq. (2.10) in~\cite{Fang:2019xat})
\begin{align}\label{eq:DeltaDelta_noLimb}
     \left.C_{\ell}^{\Delta\Delta}(z_i,z_j)\right|_{\rm lin}=& 
     \frac{2}{\pi}\int \text{d}\chi_1\, n_i(\chi_1)(1+z(\chi_1))\HH(\chi_1)\hat{b}_i(\chi_1)
     \int \text{d}\chi_2\, n_j(\chi_2)(1+z(\chi_2))\HH(\chi_2)\hat{b}_j(\chi_2)\nonumber\\
     &\times\int_0^{\infty}\text{d}k\, k^2 \frac{P_{\delta\delta}^{\rm lin}(k,z_*)}{\sigma^2_8(z_*)}j_{\ell}(k\chi_1)j_{\ell}(k\chi_2)\,,
     \quad \mbox{for $\ell< 200$}\,.
\end{align}
Eqs.~\eqref{eq:DeltaDeltaJ} and~\eqref{eq:DeltaDelta_noLimb} are similar to Eq.~\eqref{eq:DeltakappaJ}: they depend on the density power spectrum at $z_*$, on the background evolution through $\HH(z)$ and $\chi(z)$, and on the bias of the lens sample $\hat{b}_i$. Combining galaxy clustering with galaxy-galaxy lensing allows us therefore to measure separately $\hat{b}_i$ and $\hJ$. Note that in the case where GR is assumed to be correct, the measurement of $\hJ$ reduces to a measurement of $\Omega_{\rm m}(z)\sigma_8(z)$, as was proposed in~\cite{Camera:2022qfk}.

Eqs.~\eqref{eq:DeltakappaJ} and~\eqref{eq:DeltaDeltaJ} relate the angular power spectra to $\hJ$ and $\hat{b}$. In practise, however, the baseline DES analysis considered 
the angular correlation functions as observables. Here we make the same choice, to directly use the data vectors measured by the Collaboration. The angular correlation functions can be related to the angular power spectra via
\begin{align}
    w^{ij}(\theta) &= \sum_{\ell}\frac{2\ell+1}{4\pi}P_{\ell}(\cos\theta)C^{\Delta\Delta}_{\ell}(z_i,z_j)\, ,\\
    \gamma_t^{ij}(\theta) &= \sum_{\ell}\frac{2\ell+1}{4\pi\ell(\ell+1)}P_{\ell}^2(\cos\theta)C^{\Delta\kappa}_{\ell}(z_i,z_j)\,,
\end{align}
where $P_{\ell}$ and $P_{\ell}^2$ are the associated Legendre polynomials of degree $\ell$ and order 0 and 2, respectively.

The galaxy-galaxy lensing signal~\eqref{eq:DeltakappaJ} and the galaxy clustering signal~\eqref{eq:DeltaDeltaJ} depend on the boost $B(k,z)$ encoding the evolution of matter density fluctuations in the non-linear regime. Without specifying a theory of gravity, we cannot know the form of the boost. Here we follow the strategy presented in~\cite{DES:2022ygi} for extensions beyond GR, and assume that the boost is the same as in GR. We then restrict the range of angular scales
used in the analysis to minimise the impact of the boost. We consider two scenarios: one where we use all scales that were used in the standard DES analysis~\cite{DES:2021wwk}, and another one where we impose a more stringent scale cut similar to the one used in~\cite{DES:2022ygi}. These two choices are described in detail in the next section. Note that even though choosing a boost as in GR is obviously not a model-independent treatment, it does not invalidate our analysis. Suppose that we observe $\hJ$ different from GR using our formalism. This indicates without ambiguity that GR is invalid (if GR would be valid, the boost would be correct). One can then compare the measured $\hJ$ with modified gravity models and find those that are compatible with the measurements. Within these models, one can then rerun the analysis, using now the correct boost in those models to refine the constraints (which we do not expect to change by much, given that we have chosen stringent angular scale cuts). Of course, if we are in a theory of gravity where the boost is anti-correlated with $\hJ$, using a GR boost will partially hide the deviations in $\hJ$, in the regime where non-linear effects are relevant. This is unavoidable and is simply due to the fact that lensing mixes large scales and small scales leading to possible cancellations between opposite effects. Such a situation would however have a very specific signature: the measured values of $\hJ$ would deviate more and more from GR when more stringent scale cuts are used.

Once $\hJ$ has been measured, it can be compared with predictions in any theory of gravity of interest and it can be used to constrain the parameters of these theories. In particular, assuming the validity of the $\Lambda$CDM model, $\hJ$ can be used to measure $\sigma_8(z=0)$, since in this case it reduces to
\begin{align}
\label{eq:Jhat_sigma8}
\hJ(z)=\Omega_{\rm m}(z)\frac{D_1(z)}{D_1(z=0)}\sigma_8(z=0)\, .    
\end{align}

\subsection{Measurement of the Weyl potential}
\label{sec:methods}

\begin{figure}
  \includegraphics[width=0.5\linewidth]{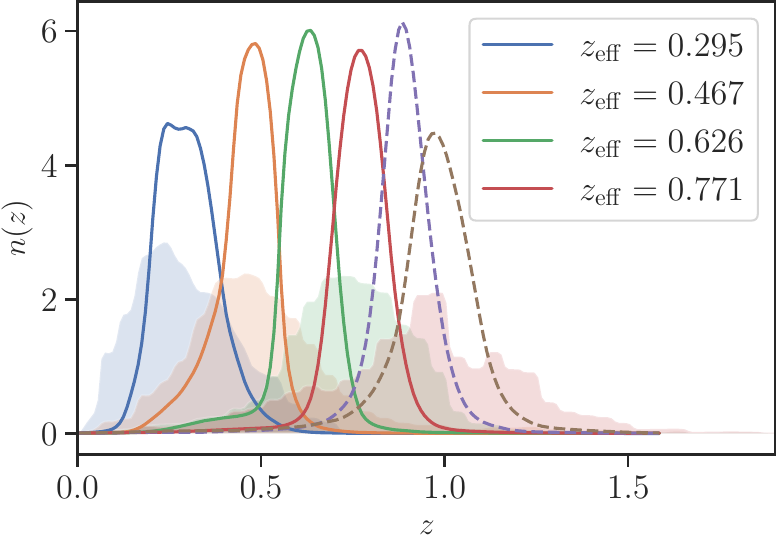}
  \caption{The solid and dashed line show the galaxy (lenses) distribution of the 6 \textsc{MagLim} tomographic bins as a function of redshift. Only the first 4 bins (solid lines) have been considered in this analysis. The effective redshift of the lenses, computed as the mean of each distribution, is also provided for information. The shaded regions show the distribution of sources. Source data are provided as a Source Data file.}
  \label{fig:nofzs}
\end{figure}

\begin{figure}
  \includegraphics[width=0.95\linewidth]{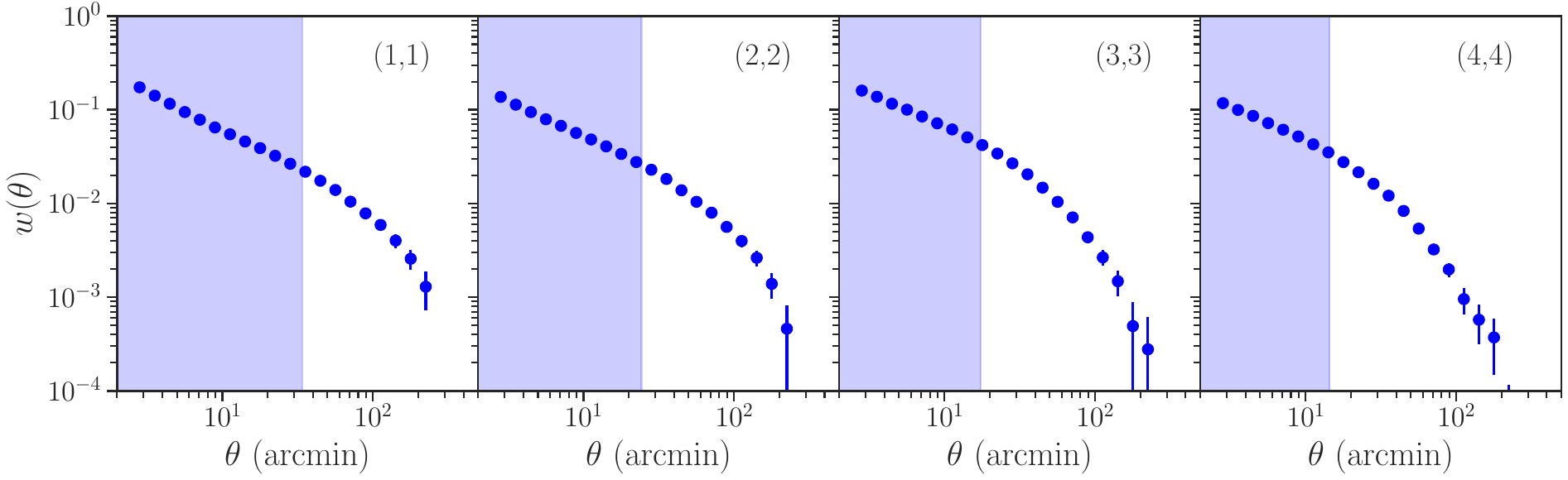}
  \caption{Measured galaxy clustering angular two-point correlation function as a function of angle for the first 4 tomographic bins of the \textsc{MagLim} sample in the DES Y3 analysis. The error bars represent 1 standard deviation. The shaded region corresponds to the measurements removed from the analysis after imposing the baseline scale cut of comoving transverse distance of 8\,Mpc$/h$. Source data are provided as a Source Data file.}
  \label{fig:wtheta}
\end{figure}

\begin{figure}
  \includegraphics[width=0.95\linewidth]{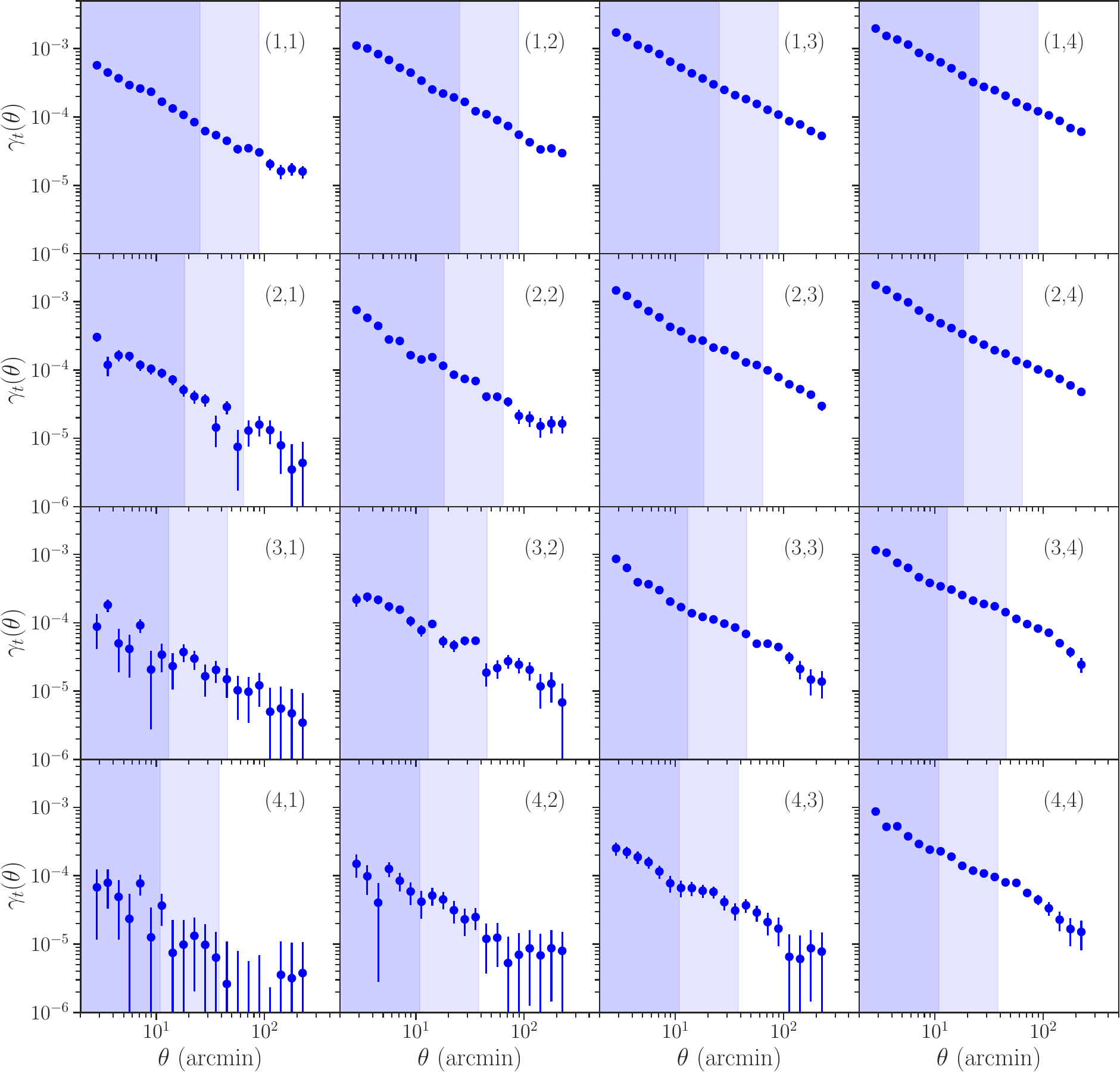}
  \caption{Measured tangential shear (galaxy-galaxy lensing signal) as a function of angle for the first 4 tomographic bins of the \textsc{MagLim} lens samples and the 4 tomographic bins of the \textsc{Metacalibration} sources in the DES Y3 analysis. Each tuple $(i,j)$ represents the combination of the $i^{\rm th}$ tomographic bin for the lenses and the $j^{\rm th}$ tomographic bin for the sources. The error bars represent 1 standard deviation. The darker shaded region corresponds to the measurements removed from the analysis after imposing the baseline scale cut of comoving transverse distance of 6\,Mpc$/h$. The lighter shaded region represents the more stringent scale cuts of 21\,Mpc$/h$ also considered in this work. Source data are provided as a Source Data file.}
  \label{fig:gammat}
\end{figure}

To measure the Weyl potential with the DES observations, we follow closely the baseline analysis for galaxy clustering and galaxy-galaxy lensing presented in~\cite{DES:2021bpo}, which is also called $2\times2$pt analysis. This includes the data sets presented in that analysis, as well as the modelling of systematic effects and the modelling of the cosmological observables. In the following we describe the main differences with respect to the baseline analysis considered in~\cite{DES:2021bpo} and
publicly-available in the \textsc{CosmoSIS} software~\cite{Zuntz:2014csq} (\href{https://cosmosis.readthedocs.io/en/latest/}{https://cosmosis.readthedocs.io/en/latest/}).

Starting with the data sets, we consider the magnitude-limited \textsc{MagLim} sample for the lenses~\cite{DES:2020ajx}, and the source sample obtained with the \textsc{Metacalibration} method~\cite{Sheldon:2017szh}. The full lens sample contains 6 tomographic bins, which would in principle allow us to determine $\hJ$ at 6 different effective redshifts. However, some residual systematic uncertainties were identified in the two bins with the highest redshift and this led to the use of only the first 4 bins in the baseline $3\times2$pt analysis~\cite{DES:2021wwk}. In this analysis we use a conservative approach and focus only on the first 4 tomographic bins, even if we only consider a $2\times2$pt analysis. The mean redshift of the first 4 tomographic \textsc{MagLim} bins are: [0.295, 0.467, 0.626, 0.771] and their distribution is plotted in Fig.\,\ref{fig:nofzs}.

In terms of modelling the (observational and instrumental) systematic effects, we include nuisance parameters for the width and the position of the distributions of the lenses, as well as nuisance parameters for the shift of the source distributions, see Supplementary Table~1 for a list of all nuisance parameters and Supplementary Discussion 4 for a discussion about degeneracies. We use the officially-provided informative priors on the corresponding nuisance parameters. We further consider the effect of intrinsic alignments in our galaxy-galaxy lensing observables. Although the baseline adopted in the $3\times2$pt analysis by the DES Collaboration was the tidal alignment and tidal torquing model~\cite{Blazek:2017wbz}, 
we follow the approach used by the Collaboration for extended models and consider the simpler non-linear alignment model~\cite{Bridle:2007ft}. 
Given the lack of understanding of intrinsic alignments in modified gravity scenarios, we prefer to take a conservative approach and treat these effects as if we were in GR. The main reason behind this choice is to avoid artificially extracting information from modified gravity from a wrong modelling of intrinsic alignments. We further marginalise over the amplitude and redshift dependence of intrinsic alignments to avoid possible biases in our posteriors due to the intrinsic alignment assumptions. 

Similar to the treatment of the intrinsic alignment of galaxies, we account for magnification effects in the lens sample assuming GR. In principle, magnification could be written in terms of the function $\hJ$, thus providing further constraints on it. However, since this effect is strongly subdominant, we expect the improvement to be negligible and we treat therefore magnification as a nuisance, that can be computed in GR without biasing the results. We fix the magnification bias parameters, as it was done in~\cite{DES:2021bpo}. We also include the effect of RSD in the galaxy clustering observable, and as for magnification we model it in GR, which again is well justified since the effect is subdominant. 
Following~\cite{DES:2021bpo} we use a linear galaxy bias model and we consider broad flat priors, given the fact that $\hat{b}_i$ now contains information relative to the growth, see Eq.\,\eqref{eq:bhat}. Finally, we account for the shear multiplicative bias using the publicly-available informative priors. Concerning the galaxy-galaxy lensing observable, we further marginalise analytically over the mass enclosed below the angular scales considered in the analysis, as was done in~\cite{DES:2021bpo}. Contrary to the baseline analysis, and in order to avoid dealing with the small scales, we prefer not to include any information from shear ratios in the likelihood.

We consider the \textsc{PolyChord}~\cite{Handley:2015fda} sampler (\href{https://github.com/PolyChord/}{https://github.com/PolyChord/}) to explore the parameter space with 500 live points, number of repetitions of 60, tolerance of 0.01, fast fraction of 0.1, and boost of the posteriors of 10, consistent with the baseline DES Y3 analysis~\cite{DES:2021wwk}.

In order to test the dependency of our results on the different assumptions, we consider four different scenarios in this analysis. More precisely, we implement two different scale cuts: a ``standard'' scale cut, where only separations above $8\,$Mpc$/h$ for galaxy clustering and $6$\,Mpc$/h$ for galaxy-galaxy lensing are included. These roughly correspond to angular scales above 34, 24, 17, and 14 arcmin, depending on the tomographic bin, for galaxy clustering, and 25, 18, 13, and 11 arcmin for galaxy-galaxy lensing. Those are the scale cuts that were used in the DES analysis~\cite{DES:2021bpo}. In addition, we also introduce a more ``pessimistic'' scale cut, where we further remove all separations below $21$\,Mpc$/h$ for galaxy-galaxy lensing (which corresponds to angular scales below 89, 64, 46, and 38 arcmin, depending on the tomographic bin). The choice of $21$\,Mpc$/h$ roughly matches the number of data points considered for galaxy-galaxy lensing in~\cite{DES:2022ygi} when testing theories of gravity beyond GR. We note that the standard scale cut for galaxy clustering is already enough to not be sensitive to the nonlinear modelling (see Table I in~\cite{DES:2022ygi}); therefore, we consider always the standard scale cut for this probe. In Figs.\,\ref{fig:wtheta}--\ref{fig:gammat} we plot the data vectors together with the two scale cuts. 

In addition to these two different scale cuts, we consider two different scenarios: one where we use $3\sigma$ Gaussian Planck priors on the cosmological parameters~\cite{Planck:2018vyg}; and another one where we do not use any information from Planck, i.e.\ we use the wide fiducial DES priors on the cosmological parameters. In this second case, the cosmological parameters will be constrained through the background evolution, the shape of the power spectrum at $z_*=10$ (which affects the shape of the $C_\ell$'s at lower redshift, see Eqs.~\eqref{eq:DeltakappaJ} and~\eqref{eq:DeltaDeltaJ}), and the contaminations (RSD, intrinsic alignment and magnification). The priors in both cases are listed in Supplementary Table~1.

To analyse the different chains and visualise posteriors, we use the \textsc{GetDist} package~\cite{Lewis:2019xzd}.
 
\subsection{Modified theories of gravity}

Since the measurements of $\hJ$ do not rely on a specific theory of gravity, they can be compared with predictions of any modified theory of gravity. The only limitation is that the theory in consideration must converge to GR at high redshift, $z_*=10$, since our measurements rely on this assumption. As illustration, we consider two cases: A) the phenomenological $\mu-\Sigma$ extensions of gravity; and B) Horndeski theories, concentrating on a few specific cases.

\subsubsection{$\mu-\Sigma$ modifications of gravity}

The phenomenological parameters $\mu$ and $\Sigma$ encode deviations from GR in the relation between the metric potentials and the matter density fluctuations,
\begin{align}
k^2 \Psi = -4 \pi \mu(a,k) G a^2 \rho \delta\,  \quad \mbox{and} \quad k^2 \Psi_W = -4 \pi \Sigma(a,k) G a^2 \rho \delta \ .   
\end{align}
In GR, $\mu=\Sigma=1$. These parameters affect $\hJ$ in the following way,
\begin{align}
\hJ(z)=\Sigma(z)\Omega_{\rm m}(z)D_1(z)\frac{\sigma_8(z_*)}{D_1(z_*)}\, , \label{eq:JmuSigma}   
\end{align}
where as before we have neglected the dependence in $k$, as is usually done in the literature, see e.g.~\cite{DES:2022ygi} (both due to the fact that current data are not constraining enough to probe a possible scale-dependence in $\mu$ and $\Sigma$, and that in many models of modified gravity the scale-dependence is negligible for sub-horizon scales). $\hJ$ depends directly on $\Sigma$, but it also depends on $\mu$ through the evolution of $D_1(z)$. Eq.~\eqref{eq:JmuSigma} underlines the difference between measuring $\hJ$, and measuring $\mu$ and $\Sigma$. $\hJ$ can be measured from lensing data on their own. The parameters $\mu$ and $\Sigma$, on the other hand, cannot be measured separately from lensing  only, since their impact on $\hJ$ is fully degenerated. To determine if $\mu$ and $\Sigma$ are consistent with GR, it is therefore necessary to include other measurements, like RSD. Moreover, in order to constrain $\mu$ from RSD, one has to assume the validity of Euler's equation for dark matter. The growth of structure is indeed not only affected by $\mu$, but also by additional forces or interactions affecting dark matter~\cite{SvevaNastassiaCamille,Bonvin:2022tii}. As a consequence, the constraints on $\Sigma$ from lensing also rely on the validity of Euler's equation. In contrast, $\hJ$ can be measured without any assumptions on the behaviour of dark matter. Finally, we see from Eq.~\eqref{eq:JmuSigma} that $\hJ$ 
depends directly on the redshift evolution of $\mu$ and $\Sigma$, which is unknown. In particular, $\hJ$ depends on $\mu(z)$ through a second order evolution equation, which requires knowledge of $\mu(z)$ not only at the redshift of the analysis, but also at all redshifts above it. 

In practice, we find that $\hJ$ is much more sensitive to $\Sigma$ than to $\mu$. This is linked to the fact that $\hJ$ is directly proportional to $\Sigma$ (see Eq.~\eqref{eq:JmuSigma}), whereas it is affected by $\mu$ only through the evolution of $D_1(z)$. Here, we aim to investigate the constraining power of $\hat{J}$, without the inclusion of additional RSD data. 
Therefore, as illustration, we choose $\mu=1$ (as in GR) and we infer $\Sigma$ from $\hJ$. 
We consider 3 different choices of time evolution, encoded through 
\begin{align}
\Sigma (z)=1+\Sigma_0g(z)\, ,
\label{eq:sigmaevol}
\end{align}
with 1) the standard evolution that has been used in the DES analysis~\cite{DES:2022ygi} and in Planck analysis~\cite{Planck:2018vyg}: $g(z)=\Omega_\Lambda(z)/\Omega_\Lambda(0)$; 2)~no evolution: $g(z)=1$ for $z\in [0,1]$ and 0 elsewhere; and 3) an exponential evolution: $g(z)=\exp(1+z)$ for $z\in [0,1]$ and 0 elsewhere. The first case is motivated by the fact that deviations from GR are linked to the accelerated expansion of the Universe, and are therefore expected to decay proportionally to the amount of dark energy. The second and third cases are not physically motivated time evolutions, but they allow us to explore the sensitivity of the data to various behaviours, drastically different from each other.

At each point of the chain, we fit for $\Sigma_0$ using Eq.~\eqref{eq:JmuSigma}, combining the four measurements of $\hJ$. The value $\sigma_8(z_*)$ is computed at each point from the cosmological parameters using \textsc{CAMB}~\cite{Lewis:1999bs}. 

\subsubsection{Horndeski theories and a non-minimal coupling of CDM}

Besides these phenomenological modifications of GR, we also consider Horndeski models~\cite{Horndeski1974}, which constitute the most general class of Lorentz-invariant scalar-tensor theories with second-order equations of motion. In the regime of linear cosmological perturbations, this broad class of theories can be expressed in an effective theory approach by a limited number of parameters~\cite{Bellini:2014fua}, including the running of the Planck mass $\alpha_{\rm M}$ and the braiding $\alpha_{\rm B}$. Following the approach of~\cite{Castello:2023zjr, Gleyzes:2015pma, Gleyzes:2015rua}, we additionally allow for dark matter to be non-minimally coupled to the scalar field, since the weak equivalence principle has not been tested for dark matter. While theoretically easier to interpret, since the modifications enter directly at the level of the Lagrangian, Horndeski models with a non-minimal coupling of CDM add more complexity compared to the phenomenological $\mu-\Sigma$ modifications. In particular, they influence the background evolution of $\HH(z)$ along with the perturbations. As previously mentioned, relaxing the assumption of the background evolution of $\Lambda$CDM would necessitate the inclusion of further data sets, which is beyond the scope of this work. Instead, for simplicity we neglect the impact on the background evolution and only constrain the parameters through their impact on $\hJ$. As such, our constraints should be seen as an illustration of the capability of this model type to explain a $\hat{J}$ measurement different from the $\Lambda$CDM prediction, rather than a full analysis of the models. 

We focus on two specific cases: 1) the case where $\alpha_{\rm B}=\alpha_{\rm M}/2$, and dark matter is minimally coupled to the scalar field, i.e.\ $\gamma_c=0$; and 2) the case where $\gamma_c\neq 0$, i.e.\ there is a non-minimal coupling of dark matter, and all other parameters are kept to their GR value. As before, we need a time evolution for the parameters, which we choose to be $g(z)=\Omega_\Lambda(z)$ as usually considered for Horndeski models. For both cases, we consider the baseline analysis, with Planck priors and standard scale cuts.

To find $\alpha_{\rm M, 0}$ and $\gamma_{\rm c, 0}$ we can use Eq.~\eqref{eq:JmuSigma}, which is also valid in Horndeski theories. For each value of the parameters, we can compute $\Sigma(z_i)=\mu(z_i)(1+\eta(z_i))$, using the relations between the model parameters $(\alpha_{\rm M}, \alpha_{\rm B}, \gamma_c)$ and the phenomenological parameters $(\mu,\eta)$ stated in Appendix~A of~\cite{Castello:2023zjr}. We then use the \textsc{EF-TIGRE} code (\href{https://github.com/Mik3M4n/EF-TIGRE}{https://github.com/Mik3M4n/EF-TIGRE}) developed in~\cite{Castello:2023zjr} to obtain $\Omega_{\rm m}(z_i)$ and $D_1(z_i)/D_1(z_\ast)$. In addition to their dependence on $\alpha_{\rm M, 0}$ and $\gamma_{\rm c, 0}$, these quantities depend also on the primordial cosmological parameters. However, since it is computationally expensive to solve the system of differential equations at each point of the chain, we solve it instead for the mean values of the primordial cosmological parameters obtained from our analysis. Similarly, we compute the mean value of $\sigma_8(z_*)$ using \textsc{CAMB}. With this, we can compute $\hJ(z_i)$ using Eq.~\eqref{eq:JmuSigma}. We then fit for $\alpha_{\rm M, 0}$, respectively for $\gamma_{c, 0}$, using the mean values for the four measured $\hat{J}$ and their covariance from our baseline analysis. Since we do not fit for $\alpha_{\rm M, 0}$ and $\gamma_{c, 0}$ at each point of the chain, we do not have posteriors for these quantities. We can, however, compute the 1$\sigma$ error bars by propagating the errors on the primordial parameters. In principle, the way in which the errors on $\Omega_{\rm m}$ and $\Omega_{\rm b}$ at initial epoch propagate into the errors on $\Omega_{\rm m}(z_i)$ and $D_1(z_i)/D_1(z_\ast)$ depends on the values of the modified gravity parameters. For simplicity, we assume that they are the same as in $\Lambda$CDM.

\section*{Data availability statement}

All the DES data used in this analysis have been produced and publicly released by the DES Collaboration~\cite{DES:2021wwk}. It is accessible through the \textsc{CosmoSIS} software~\cite{Zuntz:2014csq} in \href{https://github.com/joezuntz/cosmosis-standard-library/tree/main/likelihood/des-y3}{https://github.com/joezuntz/cosmosis-standard-library/tree/main/likelihood/des-y3}. It can also be found in \href{https://des.ncsa.illinois.edu/releases/y3a2/Y3key-products}{https://des.ncsa.illinois.edu/releases/y3a2/Y3key-products}. Source data are provided with this paper. The chains obtained in the analysis are available in~\href{https://github.com/itutusaus/cosmosis_weyl}{https://github.com/itutusaus/cosmosis{\_}weyl}\,\cite{code}.

\section*{Code availability statement}

The code used in this analysis is a modified version of \textsc{CosmoSIS} that is accessible in 
\href{https://github.com/itutusaus/cosmosis_weyl}{https://github.com/itutusaus/cosmosis{\_}weyl}\,\cite{code}. This analysis also made use of the public codes: \textsc{CAMB}~\cite{Lewis:1999bs}, \textsc{PolyChord}~\cite{Handley:2015fda}~\href{https://github.com/PolyChord/}{https://github.com/PolyChord/}, \textsc{GetDist}~\cite{Lewis:2019xzd}, and \textsc{EF-TIGRE}~\cite{Castello:2023zjr}~\href{https://github.com/Mik3M4n/EF-TIGRE}{https://github.com/Mik3M4n/EF-TIGRE}.


\acknowledgments 

We thank Ruth Durrer, Agn\`es Fert\'e, Pablo Fosalba and Ulf Leonhardt for comments and discussions. We acknowledge the use of the HPC cluster Yggdrasil at the University of Geneva for conducting our analysis. C.B. and N.G. acknowledge support from the European Research Council (ERC) under the European Union’s Horizon 2020 research and innovation program (grant agreement No. 863929; project title “Testing the law of gravity with novel large-scale structure observables”). C.B. acknowledges support from the Swiss National Science Foundation.

\section*{Author Contributions Statement}

C.B. conceived the method, developed the parameterization of the Weyl potential and led the project. I.T. modified the public code \textsc{CosmoSIS} to measure the Weyl potential and ran the different analyses on DES data. I.T., C.B. and N.G. analysed and interpreted the results. C.B. used the measurement to constrain phenomenological modifications of gravity and N.G. to constrain Horndeski theories. I.T., C.B. and N.G. wrote the manuscript. 

\section*{Competing Interests Statement}

The authors declare no competing interests.

\newpage
\section*{Supplementary information: Measurement of the Weyl potential evolution from the first three years of Dark Energy Survey data}
\appendix
\setcounter{section}{0}


\begin{supfigure}
  \includegraphics[width=0.6\linewidth]{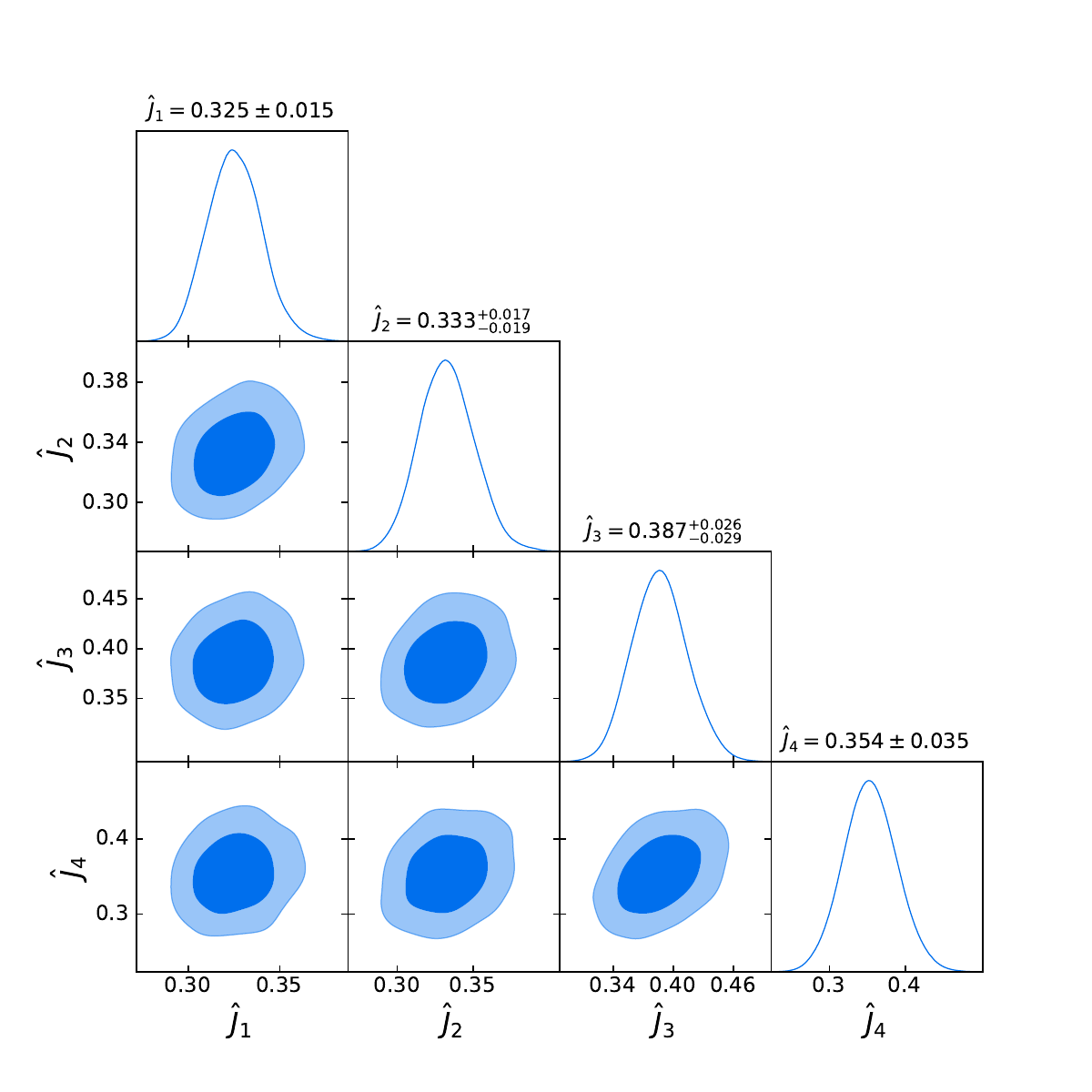}
  \caption{\justifying \textbf{Joint $\hJ_i$ constraints}. 68\% and 95\% confidence regions for the four values of $\hJ$ in the baseline scenario with Planck priors and standard scale cuts. Source data are provided as a Source Data file.}
  \label{fig:Jhat_joint}
\end{supfigure}

\begin{supfigure}[!h]
  \includegraphics[width=0.4\linewidth]{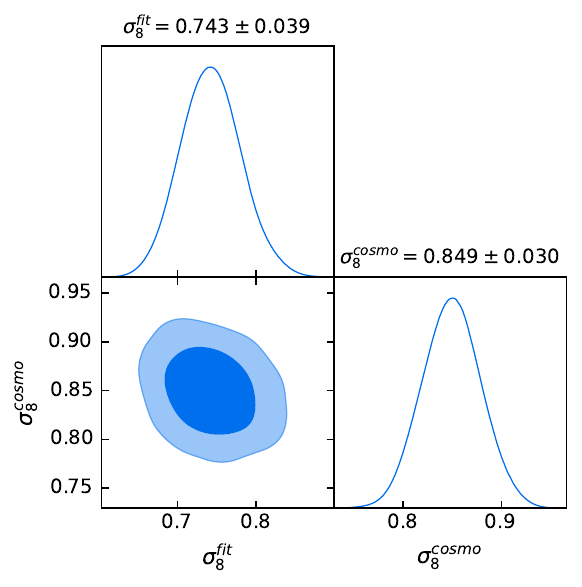}
  \caption{\justifying \textbf{Comparison of two $\sigma_8$ values}. 68\% and 95\% confidence regions for $\sigma^{\rm fit}_8(z=0)$ obtained from $\hJ$, and $\sigma_8^{\rm cosmo}(z=0)$ obtained from cosmological parameters, in the baseline scenario with Planck priors and standard scale cuts. Source data are provided as a Source Data file.}
  \label{fig:sigma8}
\end{supfigure}

\section{List of cosmological and nuisance parameters}
\label{app:parameters}

In Supplementary Table~\ref{tab:nuisance}, we list the priors, mean values and 1$\sigma$ errors that we obtain for all cosmological parameters and nuisance parameters. We show the results for the baseline analysis, i.e.\ using CMB priors and standard scale cuts, and the case where we use only DES data (also with standard scale cuts). We see that in the case with Planck priors the errors on all cosmological parameters are dominated by the priors, except for $\Omega_{\rm m}$ whose errors are a factor 2 tighter, due to constraints from the background evolution. In Supplementary Fig.~\ref{fig:cosmo}, we plot the joint constraints on cosmological parameters, in the case without Planck priors. As it can be seen in the figure, there is a weak constraint on $A_{\rm s}$ roughly equivalent to one third of the size of the prior. In order to identify the origin of such a constraint, we have rerun the analysis without any of the nuisance terms that we model in GR and without the nonlinear boost for the matter power spectrum. In this case the constraint on $A_{\rm s}$ essentially matches the size of the prior. By introducing each one of the nuisance terms at a time, we have identified that the posterior on $A_{\rm s}$ is just slightly tighter than the prior for the magnification, intrinsic alignment, and nonlinear boost contributions. Most of the constraining power to obtain the final posterior on $A_{\rm s}$ comes from the RSD contribution.

In addition to the priors listed in Supplementary Table~\ref{tab:nuisance}, we have also run an analysis with slightly different priors from Planck. Our baseline Planck priors come from the temperature and the polarisation, but without lensing, since Planck constraints from lensing assume the validity of $\Lambda$CDM at late time, which we do not want to do. However, even if lensing is not included as an additional observable, it still affects the temperature power spectrum. One way to remove this information is to consider Planck analysis with an additional parameter $A_{\rm L}$ in front of the amplitude of the lensing contribution, and to marginalise over it. This effectively removes the constraining power of lensing on cosmological parameters, ensuring that those are only constrained by early time fluctuations. As a cross-check, we have redone our baseline analysis, using these new priors from Planck. We find that the results barely change. The tension between the measured $\hJ$ and the $\Lambda$CDM prediction slightly decreases (e.g.\ from 2.81$\sigma$ to 2.5$\sigma$ in the second bin) due to the fact that Planck priors increase with this extra parameter $A_{\rm L}$. 
    
\begin{suptable}
\caption{\justifying Priors, mean values and 1$\sigma$ error bars for the cosmological parameters and nuisance parameters obtained from the analysis with and without CMB priors. The two cases are for standard scale cuts. }  \label{tab:nuisance}
  \begin{tabular}{lccccc}
  \toprule
       &  & \multicolumn{2}{c}{CMB prior} & \multicolumn{2}{c}{No CMB prior}  \\\cmidrule(r){3-4}\cmidrule(l){5-6}
     & & Parameter prior & Posterior  & Parameter prior & Posterior  \\
    \midrule
    
    Matter density & $\Omega_{\rm m}$ & $\mathcal{N}(0.3166,0.0252)$ & $0.328^{+0.011}_{-0.012}$ & $\mathcal{U}(0.1,0.9)$ & $0.294^{+0.029}_{-0.038}$\\
    
    Baryon density & $\Omega_{\rm b}$ & 
    $\mathcal{N}(0.0494116, 0.0020920)$ & $0.0489\pm 0.0021$ & $\mathcal{U}(0.03,0.07)$ & $0.0429^{+0.0048}_{-0.012}$\\
    
    Hubble parameter & $h$ & $\mathcal{N}(0.6727,0.0180)$ & $0.678\pm 0.017$ & $\mathcal{U}(0.55,0.91)$ & $0.729^{+0.091}_{-0.10}$\\
    
    Amplitude $\times 10^9$ &$A_{\rm s}$ & $\mathcal{N}(2.101, 0.102)$ & $2.13\pm 0.10$ & $\mathcal{U}(0.5,5.0)$ & $2.98^{+0.57}_{-0.69}$\\
    
    Spectral index & $n_s$ & $\mathcal{N}(0.9649, 0.0132)$ & $0.966\pm 0.013$ & $\mathcal{U}(0.87,1.07)$ & $0.981^{+0.078}_{-0.038}$\\
    
    Bias $z_1$ & $\hat{b}_1$ & $\mathcal{U}(0.1,2.0)$ & $0.957\pm 0.038$ & $\mathcal{U}(0.1,2.0)$ & $0.962\pm 0.043$ \\
    
    Bias $z_2$ & $\hat{b}_2$ & $\mathcal{U}(0.1,2.0)$ & $1.020\pm 0.041$ & $\mathcal{U}(0.1,2.0)$ & $1.054^{+0.058}_{-0.050}$ \\
    
    Bias $z_3$ & $\hat{b}_3$ & $\mathcal{U}(0.1,2.0)$ & $1.005\pm 0.033$ & $\mathcal{U}(0.1,2.0)$ & $1.037\pm 0.048$ \\
    
    Bias $z_4$ & $\hat{b}_4$ & $\mathcal{U}(0.1,2.0)$ & $0.908\pm 0.027$ & $\mathcal{U}(0.1,2.0)$ & $0.936\pm 0.041$ \\
    
    Shear calibration & $m^1$ & $\mathcal{N}(-0.0063,0.0091)$ & $-0.0059\pm 0.0091 $& $\mathcal{N}(-0.0063,0.0091)$ & $-0.0062\pm 0.0090$\\
    
    Shear calibration & $m^2$ & $\mathcal{N}(-0.0198,0.0078)$ & $-0.0205\pm 0.0078$ &$\mathcal{N}(-0.0198,0.0078)$ & $-0.0200\pm 0.0076$ \\
    
    Shear calibration & $m^3$ &$\mathcal{N}(-0.0241,0.0076)$ & $-0.0241\pm 0.0073$ & $\mathcal{N}(-0.0241,0.0076)$ & $-0.0250\pm 0.0071$\\
    
    Shear calibration & $m^4$ & $\mathcal{N}(-0.0369,0.0076)$ & $-0.0360\pm 0.0071$& $\mathcal{N}(-0.0369,0.0076)$& $-0.0363\pm 0.0071$\\
    
    Intrinsic alignment & $A_{\rm IA}$ & $\mathcal{U}(-5,5)$ & $0.227^{+0.074}_{-0.092}$ & $\mathcal{U}(-5,5)$ & $0.270\pm 0.086$\\
    
    Intrinsic alignment & $\alpha_{\rm IA}$ &$\mathcal{U}(-5,5)$ &  $-1.4^{+1.4}_{-2.8} $ &$\mathcal{U}(-5,5)$ & $-0.6^{+2.0}_{-2.3}$ \\
    
    Lens photo-$z$ shift $z_1$ & $\Delta z_l^1$ & $\mathcal{N}(-0.009,0.007)$ & $-0.0086\pm 0.0062$ & $\mathcal{N}(-0.009,0.007)$ & $-0.0088\pm 0.0061$\\
    
    Lens photo-$z$ shift $z_2$ & $\Delta z_l^2$ &$\mathcal{N}(-0.035,0.011)$ & $-0.0264\pm 0.0092$ & $\mathcal{N}(-0.035,0.011)$ & $-0.0282\pm 0.0088$\\
    
    Lens photo-$z$ shift $z_3$ & $\Delta z_l^3$ & $\mathcal{N}(-0.005,0.006)$ & $-0.0024\pm 0.0060$ & $\mathcal{N}(-0.005,0.006)$ & $-0.0018\pm 0.0058$\\
    
    Lens photo-$z$ shift $z_4$ & $\Delta z_l^4$ & $\mathcal{N}(-0.007,0.006)$ & $-0.0066\pm 0.0058$ & $\mathcal{N}(-0.007,0.006)$ & $-0.0063\pm 0.0058$\\
    
    Lens photo-$z$ stretch $z_1$ & $\sigma z_l^1$ & $\mathcal{N}(0.975,0.062)$ & $0.984\pm 0.062$ & $\mathcal{N}(0.975,0.062)$ & $0.978\pm 0.059$\\
    
    Lens photo-$z$ stretch $z_2$ & $\sigma z_l^2$ & $\mathcal{N}(1.306,0.093)$ & $1.208^{+0.078}_{-0.062}$ & $\mathcal{N}(1.306,0.093)$ & $1.236^{+0.090}_{-0.075}$\\
    
    Lens photo-$z$ stretch $z_3$ & $\sigma z_l^3$ & $\mathcal{N}(0.870,0.054)$ & $0.887\pm 0.052$ & $\mathcal{N}(0.870,0.054)$ & $0.894\pm 0.055$\\
    
    Lens photo-$z$ stretch $z_4$ & $\sigma z_l^4$ & $\mathcal{N}(0.918,0.051)$ & $0.935\pm 0.048$ & $\mathcal{N}(0.918,0.051)$ & $0.935\pm 0.050$\\
    
    Source photo-$z$ shift $z_1$ & $\Delta z_s^1$ & $\mathcal{N}(0.000,0.018)$ & $0.008\pm 0.016$ & $\mathcal{N}(0.000,0.018)$ & $0.009^{+0.015}_{-0.017}$\\
    
    Source photo-$z$ shift $z_2$ & $\Delta z_s^2$ & $\mathcal{N}(0.000,0.015)$ & $-0.0164^{+0.010}_{-0.0076}$ & $\mathcal{N}(0.000,0.015)$ & $-0.0160^{+0.0099}_{-0.0072}$\\
    
    Source photo-$z$ shift $z_3$ & $\Delta z_s^3$ & $\mathcal{N}(0.000,0.011)$ & $-0.0184\pm 0.0080$ & $\mathcal{N}(0.000,0.011)$ & $-0.0196\pm 0.0077$\\
    
    Source photo-$z$ shift $z_4$ & $\Delta z_s^4$ & $\mathcal{N}(0.000,0.017)$ & $0.006\pm 0.015$ & $\mathcal{N}(0.000,0.017)$ & $0.009\pm 0.015$\\
    \bottomrule
  \end{tabular} 
  \end{suptable}

\begin{supfigure}
  \includegraphics[width=0.85\linewidth]{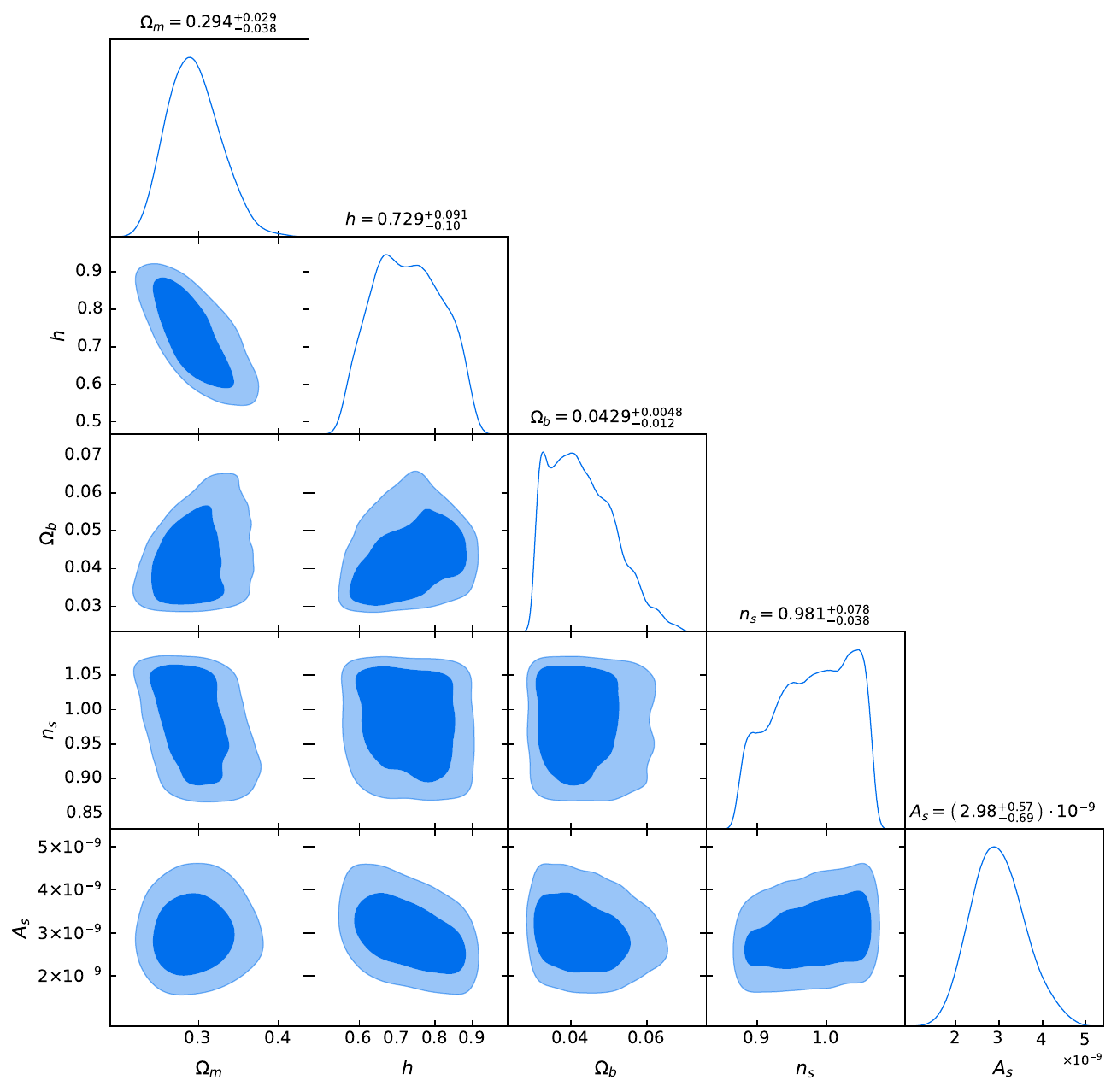}
  \caption{\justifying \textbf{Cosmological parameters constraints.} 68\% and 95\% confidence regions for the cosmological parameters in the case without Planck priors, using standard scale cuts. Source data are provided as a Source Data file.}
  \label{fig:cosmo}
\end{supfigure}

\section{Comparison with binned $\sigma_8(z)$ analysis of DES}
\label{app:binnedsigma8}

It is worth comparing our results with the binned $\sigma_8(z)$ analysis of DES~\cite{DES:2022ygi}. In this analysis, $\Lambda$CDM is assumed to be correct, and four additional free parameters, $A_i^{P_{\rm lin}}$, are added in front of the density power spectrum in the four redshift bins of the lenses. The parameter in the first bin is then fixed to 1 (to break the degeneracy with the primordial amplitude $A_{\rm s}$), whereas the other three are constrained by the data. This analysis differs from ours in the sense that it is a consistency check of $\Lambda$CDM. If the parameters are found to be different from 1 in any of the bins, then $\Lambda$CDM is inconsistent. However, this analysis does not permit a determination of the origin of the tension. 
In particular, since $\Lambda$CDM is assumed to be correct, the Weyl potential is related to the matter density fluctuation using GR. As a consequence, the angular power spectra are all sensitive to the matter power spectrum, which is multiplied by a unique parameter $A_i^{P_{\rm lin}}$ (per bin). In reality however, the clustering angular power spectrum depends on the density-density correlation; the galaxy-galaxy lensing angular power spectrum depends on the density-Weyl correlation; and the shear angular power spectrum depends on the Weyl-Weyl correlation. Modifications of gravity are not expected to modify the evolution of the density and the evolution of the Weyl potential in the same way. Therefore, encoding those into a single parameter may not be optimal. Of course, as explained in~\cite{DES:2022ygi}, the goal of such a consistency check is not to model modified gravity, but rather to assess the consistency of $\Lambda$CDM. However, the fact that one has only one parameter per bin introduces an artificial mixing between this parameter and the galaxy bias, which could in principle reduce the signature of modified gravity. Deviations from GR in the Weyl potential would indeed, in this framework, lead to modifications of the parameters $A_i^{P_{\rm lin}}$, which would artificially change the clustering power spectrum as well (since $A_i^{P_{\rm lin}}$ multiplies all spectra). If the matter density is not modified, or if it is modified in a different way than the Weyl potential, this will lead to inconsistencies in the modelling that may be minimised by artificially decreasing deviations in $A_i^{P_{\rm lin}}$ and changing the bias or other cosmological parameters. 
In contrast, with our modelling, any deviation in the Weyl potential would be automatically visible in the measurement of $\hat{J}$, while deviations in the growth of the density would affect the parameters $\hat{b}_i(z)=b_i(z)\sigma_8(z)$.

A direct comparison of our results with the binned $\sigma_8(z)$ results is not possible, since the assumptions are different. However, from Fig.~11 and 12 of~\cite{DES:2022ygi}, one can see that when using only DES data, $\sigma_8$ tends to be lower than predicted by Planck using $\Lambda$CDM, at all redshifts. Combining with Planck increases the values, bringing them back towards the $\Lambda$CDM predictions. In our case, we find that starting with a power spectrum consistent with Planck at high redshift, and letting the evolution free at low redshift, leads to a Weyl evolution below $\Lambda$CDM predictions. As for the binned $\sigma_8(z)$ analysis, this trend is not very significant. However, the main difference is that in the second bin we have a value which is 2.8$\sigma$ below the $\Lambda$CDM prediction. This is not in contradiction with the results of the binned $\sigma_8(z)$ analysis, which does not see such a feature when DES and Planck are combined. By assuming $\Lambda$CDM to be valid and fixing $A_1^{P_{\rm lin}}=1$ in the lowest bin, the constraints from different bins are not truly independent. In particular, a deviation in one bin may lead to slightly different values of the $\Lambda$CDM parameters, which in turn will artificially influence the value of $A_i^{P_{\rm lin}}$ in the other bins. Moreover, the linear galaxy bias can also reabsorb part of the deviation. In our framework this is not the case: a deviation in one redshift bin does not influence the measurements in the other bins. The correlations between $\hJ$ in different bins is purely physical, only due to the evolution of the Weyl potential. From this we see that 
identifying the quantities that are truly measured by the data helps to correctly pin-point where tensions come from. 

\section{Results using the \textsc{redMaGiC} lens sample}
\label{app:redmagic}

\begin{supfigure}
  \includegraphics[width=0.5\linewidth]{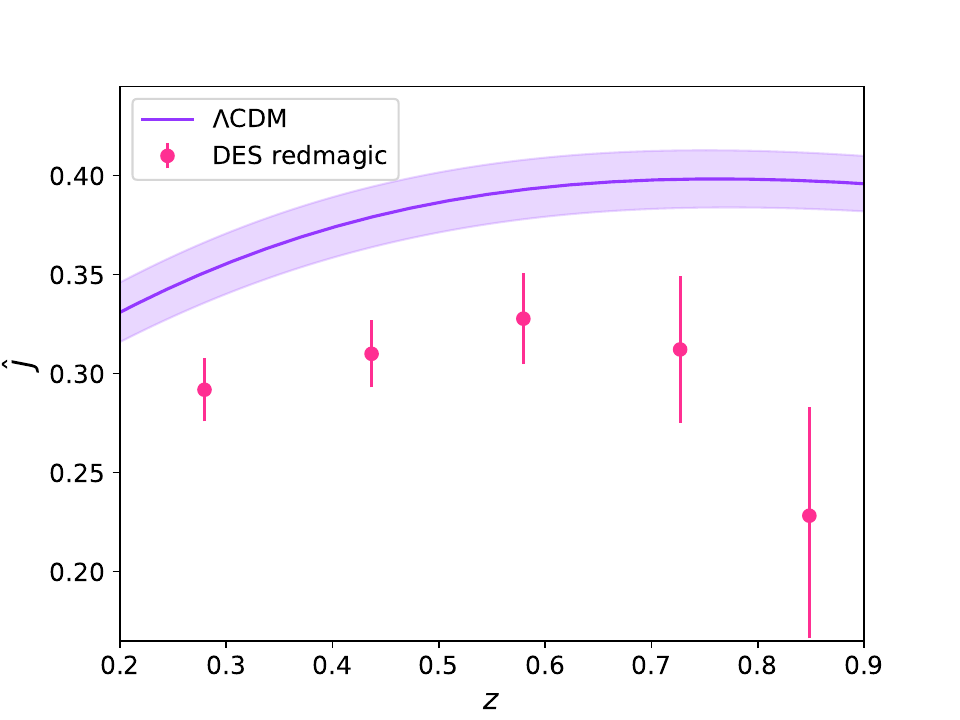}
  \caption{\justifying \textbf{\textsc{redMaGiC} results.} Measured values of $\hJ$ and 1$\sigma$ error bars at the mean redshifts of the \textsc{redMaGiC} sample, in the baseline analysis, i.e.\ using
priors from Planck on early time cosmological parameters and standard scale cuts. The purple solid line shows the prediction for $\hJ$ in $\Lambda$CDM. The shaded regions indicate the 1$\sigma$ uncertainty on this prediction due to uncertainties in the primordial cosmological parameters. Source data are provided as a Source Data file.}
  \label{fig:redmagic}
\end{supfigure}

\begin{suptable}
\caption{\justifying Mean values and 1$\sigma$ error bars for $\hat{J}(z_i)$ from the \textsc{redMaGiC} sample, in the baseline analysis.}  \label{tab:redmagic}
  \begin{tabular}{lc}
    \toprule
    $\hat{J}(z_1)$ & $0.292\pm 0.016$  \\
    $\hat{J}(z_2)$ & $0.310\pm 0.017$  \\
    $\hat{J}(z_3)$ & $0.328\pm 0.023$   \\
    $\hat{J}(z_4)$ & $0.312\pm 0.037$  \\
    $\hat{J}(z_5)$ & $0.228^{+0.055}_{-0.062}$  \\
    \bottomrule
  \end{tabular} 
\end{suptable}

In our analysis we have used the \textsc{MagLim} lens sample, which is the fiducial sample of DES. It is, however, interesting to see how the results change if we use the \textsc{redMaGiC} sample instead, which contains five redshift bins with mean redshift $z\in [0.280, 0.437, 0.580, 0.727, 0.849]$. For our baseline scenario (with Planck priors and standard scale cuts), we find that the measured $\hJ$ using \textsc{redMaGiC} are systematically below the $\Lambda$CDM predictions, as can be seen from Supplementary Fig.~\ref{fig:redmagic} and Supplementary Table~\ref{tab:redmagic}. The tension with $\Lambda$CDM in the five bins of \textsc{redMaGiC} are $[2.7, 3, 2.4, 2.2, 3]\sigma$. Starting from these measurements and assuming the validity of $\Lambda$CDM, we find that the inferred value for $\sigma_8$ at late time is $\sigma^{\rm fit}_8(z=0)=0.626^{+0.038}_{-0.043}$, which is 4.1$\sigma$ below the value inferred from early time cosmological parameters: $\sigma^{\rm cosmo}_8(z=0)=0.819\pm 0.028$. 

This non-negligible tension with respect to $\Lambda$CDM predictions is in agreement with the results of the DES Collaboration, who found that $\sigma_8$ measured from galaxy-galaxy lensing and galaxy clustering is significantly lower using \textsc{redMaGiC} than \textsc{MagLim}~\cite{DES:2021zxv}. To account for this, an artificial parameter $X_{\rm lens}$ was introduced, encoding a difference in the galaxy bias obtained from galaxy clustering and the bias obtained from galaxy-galaxy lensing. This parameter was found to be in 5$\sigma$ tension with its expected value of 1, reflecting the fact that the galaxy clustering amplitude was systematically larger than the galaxy-galaxy lensing amplitude~\cite{DES:2021zxv,DES:2021wwk}. This is fully in line with our findings: leaving the function $\hJ$ free, we find that the data prefer a low value of $\hJ$ consistent with a galaxy-galaxy lensing correlation lower than in $\Lambda$CDM. The DES Collaboration relates the low value of $X_{\rm lens}$ to systematic effects due to the selection of galaxies in the \textsc{redMaGiC} sample. Such systematic effects would bias our measurements of $\hJ$, which is degenerated with the galaxy bias in the galaxy-galaxy lensing signal. One way to investigate this would be to measure $\hJ$ from shear-shear correlations and see if the results are consistent. This would however require to modify our analysis, by parameterizing $\hJ$ along the line-of-sight (e.g.\ with piece-wise constant values in chosen redshift bins). We defer this to a future analysis.

\section{Degeneracy with nuisance parameters}
\label{app:plots}

In Supplementary Fig.~\ref{fig:shift} we show the joint distributions, in the second redshift bin, of $\hJ$ and the two nuisance parameters $\Delta z_l$ (shift) and $\sigma z_l$ (width) that govern the photometric redshift distribution of the lenses, in the case with standard scale cuts and no Planck prior. From this plot we see that $\hJ$ is degenerated with the stretch of the photometric redshift distribution of the lens sample. This stretch is added as a free parameter in the baseline DES analysis, together with a shift, to account for photometric redshift systematic effects~\cite{DES:2021bpo}. The stretch and the shift are calibrated using clustering with spectroscopic surveys~\cite{DES:2020sjz}, leading to tight priors on these parameters. We find that a large value of the stretch, i.e.\ a distribution which is more spread, can be counterbalanced by smaller values of $\hJ$, and vice-versa. This can be understood from Eq.~(3): increasing the width of $n_i(z)$ tends to increase the values of the integral over $z$, due to the lensing kernel $(\chi(z)-\chi'(z'))/(\chi(z)\chi'(z'))$, which increases when $\chi(z)$ is allowed to take smaller values. In order to have $\hJ$ in 1$\sigma$ agreement with $\Lambda$CDM predictions in the second bin, we would need a stretch of 0.94, which is in clear tension with the values obtained from calibration: $1.306\pm 0.093$ (the mean value from our analysis is $1.236^{+0.090}_{-0.075}$. Supplementary Table~\ref{tab:nuisance} in~\ref{app:parameters} summarises the mean values and error bars for all nuisance parameters.

\begin{supfigure}
  \includegraphics[width=0.5\linewidth]{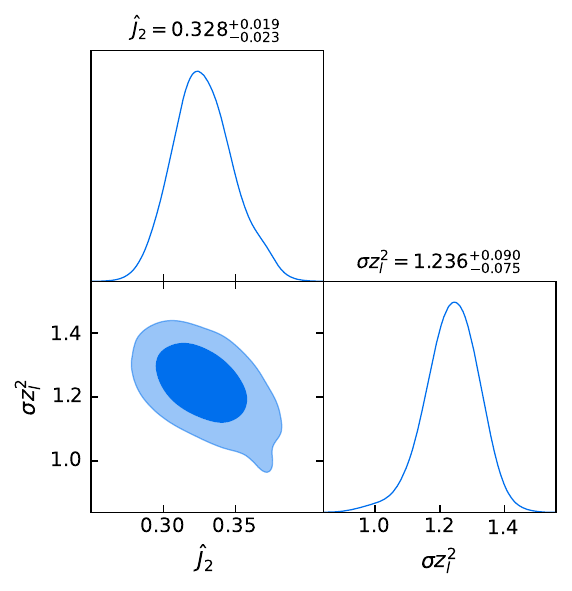}
  \caption{\justifying \textbf{Degeneracy with photometric distribution.} 68\% and 95\% confidence regions for $\hJ$ and the stretch $\sigma z_l$ of the photometric distribution in the second bin of the \textsc{MagLim} sample, where the tension is the largest. Here, we show the case without Planck prior and with standard scale cuts. Source data are provided as a Source Data file.}
  \label{fig:shift}
\end{supfigure}

\clearpage

\bibliographystyle{Bonvinetal}
\bibliography{jdes}

\ifx\mcitethebibliography\mciteundefinedmacro
\PackageError{LHCb.bst}{mciteplus.sty has not been loaded}
{This bibstyle requires the use of the mciteplus package.}\fi
\providecommand{\href}[2]{#2}
\begin{mcitethebibliography}{10}
\mciteSetBstSublistMode{n}
\mciteSetBstMaxWidthForm{subitem}{\alph{mcitesubitemcount})}
\mciteSetBstSublistLabelBeginEnd{\mcitemaxwidthsubitemform\space}
{\relax}{\relax}

\bibitem{SupernovaSearchTeam:1998fmf}
Supernova Search Team, A.~G. Riess, A.~V. Filippenko, P.~Challis,
  A.~Clocchiatti, A.~Diercks {\em et~al.},
  \ifthenelse{\boolean{articletitles}}{\emph{{Observational evidence from
  supernovae for an accelerating universe and a cosmological constant}},
  }{}\href{https://doi.org/10.1086/300499}{\aj \textbf{116} (1998) 1009},
  \href{http://arxiv.org/abs/astro-ph/9805201}{{\normalfont\ttfamily
  arXiv:astro-ph/9805201}}\relax
\mciteBstWouldAddEndPuncttrue
\mciteSetBstMidEndSepPunct{\mcitedefaultmidpunct}
{\mcitedefaultendpunct}{\mcitedefaultseppunct}\relax
\EndOfBibitem
\bibitem{SupernovaCosmologyProject:1998vns}
Supernova Cosmology Project, S.~Perlmutter {\em et~al.},
  \ifthenelse{\boolean{articletitles}}{\emph{{Measurements of $\Omega$ and
  $\Lambda$ from 42 high redshift supernovae}},
  }{}\href{https://doi.org/10.1086/307221}{Astrophys.\ J.\  \textbf{517} (1999)
  565}, \href{http://arxiv.org/abs/astro-ph/9812133}{{\normalfont\ttfamily
  arXiv:astro-ph/9812133}}\relax
\mciteBstWouldAddEndPuncttrue
\mciteSetBstMidEndSepPunct{\mcitedefaultmidpunct}
{\mcitedefaultendpunct}{\mcitedefaultseppunct}\relax
\EndOfBibitem
\bibitem{Horndeski1974}
G.~W. Horndeski, \ifthenelse{\boolean{articletitles}}{\emph{Second-order
  scalar-tensor field equations in a four-dimensional space},
  }{}\href{https://doi.org/10.1007/BF01807638}{International Journal of
  Theoretical Physics \textbf{10} (1974) 363}\relax
\mciteBstWouldAddEndPuncttrue
\mciteSetBstMidEndSepPunct{\mcitedefaultmidpunct}
{\mcitedefaultendpunct}{\mcitedefaultseppunct}\relax
\EndOfBibitem
\bibitem{Noller:2018wyv}
J.~Noller and A.~Nicola,
  \ifthenelse{\boolean{articletitles}}{\emph{{Cosmological parameter
  constraints for Horndeski scalar-tensor gravity}},
  }{}\href{https://doi.org/10.1103/PhysRevD.99.103502}{Phys.\ Rev.\ D
  \textbf{99} (2019) 103502},
  \href{http://arxiv.org/abs/1811.12928}{{\normalfont\ttfamily
  arXiv:1811.12928}}\relax
\mciteBstWouldAddEndPuncttrue
\mciteSetBstMidEndSepPunct{\mcitedefaultmidpunct}
{\mcitedefaultendpunct}{\mcitedefaultseppunct}\relax
\EndOfBibitem
\bibitem{Koyama:2015vza}
K.~Koyama, \ifthenelse{\boolean{articletitles}}{\emph{{Cosmological Tests of
  Modified Gravity}},
  }{}\href{https://doi.org/10.1088/0034-4885/79/4/046902}{Rept.\ Prog.\ Phys.\
  \textbf{79} (2016) 046902},
  \href{http://arxiv.org/abs/1504.04623}{{\normalfont\ttfamily
  arXiv:1504.04623}}\relax
\mciteBstWouldAddEndPuncttrue
\mciteSetBstMidEndSepPunct{\mcitedefaultmidpunct}
{\mcitedefaultendpunct}{\mcitedefaultseppunct}\relax
\EndOfBibitem
\bibitem{DES:2022ygi}
DES collaboration, T.~M.~C. Abbott, M.~Aguena, A.~Alarcon, O.~Alves, A.~Amon
  {\em et~al.}, \ifthenelse{\boolean{articletitles}}{\emph{{Dark Energy Survey
  Year 3 Results: Constraints on extensions to $\Lambda${CDM} with weak lensing
  and galaxy clustering}},
  }{}\href{https://doi.org/10.1103/PhysRevD.107.083504}{\prd \textbf{107}
  (2023) 083504}, \href{http://arxiv.org/abs/2207.05766}{{\normalfont\ttfamily
  arXiv:2207.05766}}\relax
\mciteBstWouldAddEndPuncttrue
\mciteSetBstMidEndSepPunct{\mcitedefaultmidpunct}
{\mcitedefaultendpunct}{\mcitedefaultseppunct}\relax
\EndOfBibitem
\bibitem{Pogosian:2021mcs}
L.~Pogosian, M.~Raveri, K.~Koyama, M.~Martinelli, A.~Silvestri {\em et~al.},
  \ifthenelse{\boolean{articletitles}}{\emph{{Imprints of cosmological tensions
  in reconstructed gravity}},
  }{}\href{https://doi.org/10.1038/s41550-022-01808-7}{Nature Astron.\
  \textbf{6} (2022) 1484},
  \href{http://arxiv.org/abs/2107.12992}{{\normalfont\ttfamily
  arXiv:2107.12992}}\relax
\mciteBstWouldAddEndPuncttrue
\mciteSetBstMidEndSepPunct{\mcitedefaultmidpunct}
{\mcitedefaultendpunct}{\mcitedefaultseppunct}\relax
\EndOfBibitem
\bibitem{SvevaNastassiaCamille}
S.~Castello, N.~Grimm, and C.~Bonvin,
  \ifthenelse{\boolean{articletitles}}{\emph{{Rescuing constraints on modified
  gravity using gravitational redshift in large-scale structure}},
  }{}\href{https://doi.org/10.1103/PhysRevD.106.083511}{\prd \textbf{106}
  (2022) 083511}, \href{http://arxiv.org/abs/2204.11507}{{\normalfont\ttfamily
  arXiv:2204.11507}}\relax
\mciteBstWouldAddEndPuncttrue
\mciteSetBstMidEndSepPunct{\mcitedefaultmidpunct}
{\mcitedefaultendpunct}{\mcitedefaultseppunct}\relax
\EndOfBibitem
\bibitem{Bonvin:2022tii}
C.~Bonvin and L.~Pogosian, \ifthenelse{\boolean{articletitles}}{\emph{{Modified
  Einstein versus Modified Euler for Dark Matter}},
  }{}\href{https://doi.org/10.1038/s41550-023-02003-y}{Nature Astron.\
  \textbf{7} (2023) 1127},
  \href{http://arxiv.org/abs/2209.03614}{{\normalfont\ttfamily
  arXiv:2209.03614}}\relax
\mciteBstWouldAddEndPuncttrue
\mciteSetBstMidEndSepPunct{\mcitedefaultmidpunct}
{\mcitedefaultendpunct}{\mcitedefaultseppunct}\relax
\EndOfBibitem
\bibitem{Song:2008qt}
Y.-S. Song and W.~J. Percival,
  \ifthenelse{\boolean{articletitles}}{\emph{{Reconstructing the history of
  structure formation using Redshift Distortions}},
  }{}\href{https://doi.org/10.1088/1475-7516/2009/10/004}{JCAP \textbf{10}
  (2009) 004}, \href{http://arxiv.org/abs/0807.0810}{{\normalfont\ttfamily
  arXiv:0807.0810}}\relax
\mciteBstWouldAddEndPuncttrue
\mciteSetBstMidEndSepPunct{\mcitedefaultmidpunct}
{\mcitedefaultendpunct}{\mcitedefaultseppunct}\relax
\EndOfBibitem
\bibitem{Blake_2011}
C.~Blake, S.~Brough, M.~Colless, C.~Contreras, W.~Couch {\em et~al.},
  \ifthenelse{\boolean{articletitles}}{\emph{The {WiggleZ} dark energy survey:
  the growth rate of cosmic structure since redshift z=0.9},
  }{}\href{https://doi.org/10.1111/j.1365-2966.2011.18903.x}{\mnras
  \textbf{415} (2011) 2876},
  \href{http://arxiv.org/abs/1104.2948}{{\normalfont\ttfamily
  arXiv:1104.2948}}\relax
\mciteBstWouldAddEndPuncttrue
\mciteSetBstMidEndSepPunct{\mcitedefaultmidpunct}
{\mcitedefaultendpunct}{\mcitedefaultseppunct}\relax
\EndOfBibitem
\bibitem{eBOSS:2020yzd}
eBOSS collaboration, S.~Alam, M.~Aubert, S.~Avila, C.~Balland, J.~E. Bautista
  {\em et~al.}, \ifthenelse{\boolean{articletitles}}{\emph{{Completed {SDSS-IV}
  extended Baryon Oscillation Spectroscopic Survey: Cosmological implications
  from two decades of spectroscopic surveys at the Apache Point Observatory}},
  }{}\href{https://doi.org/10.1103/PhysRevD.103.083533}{\prd \textbf{103}
  (2021) 083533}, \href{http://arxiv.org/abs/2007.08991}{{\normalfont\ttfamily
  arXiv:2007.08991}}\relax
\mciteBstWouldAddEndPuncttrue
\mciteSetBstMidEndSepPunct{\mcitedefaultmidpunct}
{\mcitedefaultendpunct}{\mcitedefaultseppunct}\relax
\EndOfBibitem
\bibitem{Tutusaus:2022cab}
I.~Tutusaus, D.~Sobral-Blanco, and C.~Bonvin,
  \ifthenelse{\boolean{articletitles}}{\emph{{Combining gravitational lensing
  and gravitational redshift to measure the anisotropic stress with future
  galaxy surveys}}, }{}\href{https://doi.org/10.1103/physrevd.107.083526}{\prd
  \textbf{107} (2023) },
  \href{http://arxiv.org/abs/2209.08987}{{\normalfont\ttfamily
  arXiv:2209.08987}}\relax
\mciteBstWouldAddEndPuncttrue
\mciteSetBstMidEndSepPunct{\mcitedefaultmidpunct}
{\mcitedefaultendpunct}{\mcitedefaultseppunct}\relax
\EndOfBibitem
\bibitem{DES:2021wwk}
DES collaboration, T.~M.~C. {Abbott}, M.~{Aguena}, A.~{Alarcon}, S.~{Allam},
  O.~{Alves} {\em et~al.}, \ifthenelse{\boolean{articletitles}}{\emph{{Dark
  Energy Survey Year 3 results: Cosmological constraints from galaxy clustering
  and weak lensing}},
  }{}\href{https://doi.org/10.1103/PhysRevD.105.023520}{\prd \textbf{105}
  (2022) 023520}, \href{http://arxiv.org/abs/2105.13549}{{\normalfont\ttfamily
  arXiv:2105.13549}}\relax
\mciteBstWouldAddEndPuncttrue
\mciteSetBstMidEndSepPunct{\mcitedefaultmidpunct}
{\mcitedefaultendpunct}{\mcitedefaultseppunct}\relax
\EndOfBibitem
\bibitem{Planck:2018vyg}
Planck, N.~Aghanim, Y.~Akrami, M.~Ashdown, J.~Aumont, C.~Baccigalupi {\em
  et~al.}, \ifthenelse{\boolean{articletitles}}{\emph{{Planck 2018 results. VI.
  Cosmological parameters}},
  }{}\href{https://doi.org/10.1051/0004-6361/201833910}{\aap \textbf{641}
  (2020) A6}, \href{http://arxiv.org/abs/1807.06209}{{\normalfont\ttfamily
  arXiv:1807.06209}}, [Erratum: \aap 652, C4 (2021)]\relax
\mciteBstWouldAddEndPuncttrue
\mciteSetBstMidEndSepPunct{\mcitedefaultmidpunct}
{\mcitedefaultendpunct}{\mcitedefaultseppunct}\relax
\EndOfBibitem
\bibitem{Garcia-Garcia:2021unp}
C.~Garc\'\i{}a-Garc\'\i{}a, J.~R. Zapatero, D.~Alonso, E.~Bellini, P.~G.
  Ferreira {\em et~al.}, \ifthenelse{\boolean{articletitles}}{\emph{{The growth
  of density perturbations in the last \ensuremath{\sim}10 billion years from
  tomographic large-scale structure data}},
  }{}\href{https://doi.org/10.1088/1475-7516/2021/10/030}{JCAP \textbf{10}
  (2021) 030}, \href{http://arxiv.org/abs/2105.12108}{{\normalfont\ttfamily
  arXiv:2105.12108}}\relax
\mciteBstWouldAddEndPuncttrue
\mciteSetBstMidEndSepPunct{\mcitedefaultmidpunct}
{\mcitedefaultendpunct}{\mcitedefaultseppunct}\relax
\EndOfBibitem
\bibitem{DES:2015pcw}
DES, E.~Rozo {\em et~al.},
  \ifthenelse{\boolean{articletitles}}{\emph{{redMaGiC: Selecting Luminous Red
  Galaxies from the DES Science Verification Data}},
  }{}\href{https://doi.org/10.1093/mnras/stw1281}{Mon.\ Not.\ Roy.\ Astron.\
  Soc.\  \textbf{461} (2016) 1431},
  \href{http://arxiv.org/abs/1507.05460}{{\normalfont\ttfamily
  arXiv:1507.05460}}\relax
\mciteBstWouldAddEndPuncttrue
\mciteSetBstMidEndSepPunct{\mcitedefaultmidpunct}
{\mcitedefaultendpunct}{\mcitedefaultseppunct}\relax
\EndOfBibitem
\bibitem{ACT:2023ipp}
ACT, DES, G.~A. Marques {\em et~al.},
  \ifthenelse{\boolean{articletitles}}{\emph{{Cosmological constraints from the
  tomography of DES-Y3 galaxies with CMB lensing from ACT DR4}},
  }{}\href{https://doi.org/10.1088/1475-7516/2024/01/033}{JCAP \textbf{01}
  (2024) 033}, \href{http://arxiv.org/abs/2306.17268}{{\normalfont\ttfamily
  arXiv:2306.17268}}\relax
\mciteBstWouldAddEndPuncttrue
\mciteSetBstMidEndSepPunct{\mcitedefaultmidpunct}
{\mcitedefaultendpunct}{\mcitedefaultseppunct}\relax
\EndOfBibitem
\bibitem{Chen:2022jzq}
S.-F. Chen, M.~White, J.~DeRose, and N.~Kokron,
  \ifthenelse{\boolean{articletitles}}{\emph{{Cosmological analysis of
  three-dimensional BOSS galaxy clustering and Planck CMB lensing cross
  correlations via Lagrangian perturbation theory}},
  }{}\href{https://doi.org/10.1088/1475-7516/2022/07/041}{JCAP \textbf{07}
  (2022) 041}, \href{http://arxiv.org/abs/2204.10392}{{\normalfont\ttfamily
  arXiv:2204.10392}}\relax
\mciteBstWouldAddEndPuncttrue
\mciteSetBstMidEndSepPunct{\mcitedefaultmidpunct}
{\mcitedefaultendpunct}{\mcitedefaultseppunct}\relax
\EndOfBibitem
\bibitem{White:2021yvw}
M.~White {\em et~al.}, \ifthenelse{\boolean{articletitles}}{\emph{{Cosmological
  constraints from the tomographic cross-correlation of DESI Luminous Red
  Galaxies and Planck CMB lensing}},
  }{}\href{https://doi.org/10.1088/1475-7516/2022/02/007}{JCAP \textbf{02}
  (2022) 007}, \href{http://arxiv.org/abs/2111.09898}{{\normalfont\ttfamily
  arXiv:2111.09898}}\relax
\mciteBstWouldAddEndPuncttrue
\mciteSetBstMidEndSepPunct{\mcitedefaultmidpunct}
{\mcitedefaultendpunct}{\mcitedefaultseppunct}\relax
\EndOfBibitem
\bibitem{Alonso:2023guh}
D.~Alonso, G.~Fabbian, K.~Storey-Fisher, A.-C. Eilers,
  C.~Garc\'\i{}a-Garc\'\i{}a {\em et~al.},
  \ifthenelse{\boolean{articletitles}}{\emph{{Constraining cosmology with the
  Gaia-unWISE Quasar Catalog and CMB lensing: structure growth}},
  }{}\href{https://doi.org/10.1088/1475-7516/2023/11/043}{JCAP \textbf{11}
  (2023) 043}, \href{http://arxiv.org/abs/2306.17748}{{\normalfont\ttfamily
  arXiv:2306.17748}}\relax
\mciteBstWouldAddEndPuncttrue
\mciteSetBstMidEndSepPunct{\mcitedefaultmidpunct}
{\mcitedefaultendpunct}{\mcitedefaultseppunct}\relax
\EndOfBibitem
\bibitem{Piccirilli:2024xgo}
G.~Piccirilli, G.~Fabbian, D.~Alonso, K.~Storey-Fisher, J.~Carron {\em et~al.},
  \ifthenelse{\boolean{articletitles}}{\emph{{Growth history and quasar bias
  evolution at z \ensuremath{<} 3 from Quaia}},
  }{}\href{http://arxiv.org/abs/2402.05761}{{\normalfont\ttfamily
  arXiv:2402.05761}}\relax
\mciteBstWouldAddEndPuncttrue
\mciteSetBstMidEndSepPunct{\mcitedefaultmidpunct}
{\mcitedefaultendpunct}{\mcitedefaultseppunct}\relax
\EndOfBibitem
\bibitem{ACT:2023kun}
ACT, M.~S. Madhavacheril {\em et~al.},
  \ifthenelse{\boolean{articletitles}}{\emph{{The Atacama Cosmology Telescope:
  DR6 Gravitational Lensing Map and Cosmological Parameters}},
  }{}\href{https://doi.org/10.3847/1538-4357/acff5f}{Astrophys.\ J.\
  \textbf{962} (2024) 113},
  \href{http://arxiv.org/abs/2304.05203}{{\normalfont\ttfamily
  arXiv:2304.05203}}\relax
\mciteBstWouldAddEndPuncttrue
\mciteSetBstMidEndSepPunct{\mcitedefaultmidpunct}
{\mcitedefaultendpunct}{\mcitedefaultseppunct}\relax
\EndOfBibitem
\bibitem{Esposito:2022plo}
M.~Esposito, V.~Ir\v{s}i\v{c}, M.~Costanzi, S.~Borgani, A.~Saro {\em et~al.},
  \ifthenelse{\boolean{articletitles}}{\emph{{Weighing cosmic structures with
  clusters of galaxies and the intergalactic medium}},
  }{}\href{https://doi.org/10.1093/mnras/stac1825}{Mon.\ Not.\ Roy.\ Astron.\
  Soc.\  \textbf{515} (2022) 857},
  \href{http://arxiv.org/abs/2202.00974}{{\normalfont\ttfamily
  arXiv:2202.00974}}\relax
\mciteBstWouldAddEndPuncttrue
\mciteSetBstMidEndSepPunct{\mcitedefaultmidpunct}
{\mcitedefaultendpunct}{\mcitedefaultseppunct}\relax
\EndOfBibitem
\bibitem{Grimm:2024fui}
N.~Grimm, C.~Bonvin, and I.~Tutusaus,
  \ifthenelse{\boolean{articletitles}}{\emph{{New measurements of $E_G$:
  Testing General Relativity with the Weyl potential and galaxy velocities}},
  }{}\href{http://arxiv.org/abs/2403.13709}{{\normalfont\ttfamily
  arXiv:2403.13709}}\relax
\mciteBstWouldAddEndPuncttrue
\mciteSetBstMidEndSepPunct{\mcitedefaultmidpunct}
{\mcitedefaultendpunct}{\mcitedefaultseppunct}\relax
\EndOfBibitem
\bibitem{Castello:2023zjr}
S.~Castello, M.~Mancarella, N.~Grimm, D.~Sobral-Blanco, I.~Tutusaus {\em
  et~al.}, \ifthenelse{\boolean{articletitles}}{\emph{{Gravitational redshift
  constraints on the effective theory of interacting dark energy}},
  }{}\href{https://doi.org/10.1088/1475-7516/2024/05/003}{JCAP \textbf{05}
  (2024) 003}, \href{http://arxiv.org/abs/2311.14425}{{\normalfont\ttfamily
  arXiv:2311.14425}}\relax
\mciteBstWouldAddEndPuncttrue
\mciteSetBstMidEndSepPunct{\mcitedefaultmidpunct}
{\mcitedefaultendpunct}{\mcitedefaultseppunct}\relax
\EndOfBibitem
\bibitem{Brans:1961sx}
C.~Brans and R.~H. Dicke, \ifthenelse{\boolean{articletitles}}{\emph{{Mach's
  principle and a relativistic theory of gravitation}},
  }{}\href{https://doi.org/10.1103/PhysRev.124.925}{Phys.\ Rev.\  \textbf{124}
  (1961) 925}\relax
\mciteBstWouldAddEndPuncttrue
\mciteSetBstMidEndSepPunct{\mcitedefaultmidpunct}
{\mcitedefaultendpunct}{\mcitedefaultseppunct}\relax
\EndOfBibitem
\bibitem{Sawicki:2012re}
I.~Sawicki, I.~D. Saltas, L.~Amendola, and M.~Kunz,
  \ifthenelse{\boolean{articletitles}}{\emph{{Consistent perturbations in an
  imperfect fluid}},
  }{}\href{https://doi.org/10.1088/1475-7516/2013/01/004}{JCAP \textbf{01}
  (2013) 004}, \href{http://arxiv.org/abs/1208.4855}{{\normalfont\ttfamily
  arXiv:1208.4855}}\relax
\mciteBstWouldAddEndPuncttrue
\mciteSetBstMidEndSepPunct{\mcitedefaultmidpunct}
{\mcitedefaultendpunct}{\mcitedefaultseppunct}\relax
\EndOfBibitem
\bibitem{Carroll:2003wy}
S.~M. Carroll, V.~Duvvuri, M.~Trodden, and M.~S. Turner,
  \ifthenelse{\boolean{articletitles}}{\emph{{Is cosmic speed - up due to new
  gravitational physics?}},
  }{}\href{https://doi.org/10.1103/PhysRevD.70.043528}{Phys.\ Rev.\ D
  \textbf{70} (2004) 043528},
  \href{http://arxiv.org/abs/astro-ph/0306438}{{\normalfont\ttfamily
  arXiv:astro-ph/0306438}}\relax
\mciteBstWouldAddEndPuncttrue
\mciteSetBstMidEndSepPunct{\mcitedefaultmidpunct}
{\mcitedefaultendpunct}{\mcitedefaultseppunct}\relax
\EndOfBibitem
\bibitem{Song:2006ej}
Y.-S. Song, W.~Hu, and I.~Sawicki,
  \ifthenelse{\boolean{articletitles}}{\emph{{The Large Scale Structure of f(R)
  Gravity}}, }{}\href{https://doi.org/10.1103/PhysRevD.75.044004}{Phys.\ Rev.\
  D \textbf{75} (2007) 044004},
  \href{http://arxiv.org/abs/astro-ph/0610532}{{\normalfont\ttfamily
  arXiv:astro-ph/0610532}}\relax
\mciteBstWouldAddEndPuncttrue
\mciteSetBstMidEndSepPunct{\mcitedefaultmidpunct}
{\mcitedefaultendpunct}{\mcitedefaultseppunct}\relax
\EndOfBibitem
\bibitem{Carroll:2006jn}
S.~M. Carroll, I.~Sawicki, A.~Silvestri, and M.~Trodden,
  \ifthenelse{\boolean{articletitles}}{\emph{{Modified-Source Gravity and
  Cosmological Structure Formation}},
  }{}\href{https://doi.org/10.1088/1367-2630/8/12/323}{New J.\ Phys.\
  \textbf{8} (2006) 323},
  \href{http://arxiv.org/abs/astro-ph/0607458}{{\normalfont\ttfamily
  arXiv:astro-ph/0607458}}\relax
\mciteBstWouldAddEndPuncttrue
\mciteSetBstMidEndSepPunct{\mcitedefaultmidpunct}
{\mcitedefaultendpunct}{\mcitedefaultseppunct}\relax
\EndOfBibitem
\bibitem{Vollick:2003aw}
D.~N. Vollick, \ifthenelse{\boolean{articletitles}}{\emph{{1/R Curvature
  corrections as the source of the cosmological acceleration}},
  }{}\href{https://doi.org/10.1103/PhysRevD.68.063510}{Phys.\ Rev.\ D
  \textbf{68} (2003) 063510},
  \href{http://arxiv.org/abs/astro-ph/0306630}{{\normalfont\ttfamily
  arXiv:astro-ph/0306630}}\relax
\mciteBstWouldAddEndPuncttrue
\mciteSetBstMidEndSepPunct{\mcitedefaultmidpunct}
{\mcitedefaultendpunct}{\mcitedefaultseppunct}\relax
\EndOfBibitem
\bibitem{Kilo-DegreeSurvey:2023gfr}
Kilo-Degree Survey, Dark Energy Survey, T.~M.~C. Abbott {\em et~al.},
  \ifthenelse{\boolean{articletitles}}{\emph{{DES Y3 + KiDS-1000: Consistent
  cosmology combining cosmic shear surveys}},
  }{}\href{https://doi.org/10.21105/astro.2305.17173}{Open J.\ Astrophys.\
  \textbf{6} (2023) 2305.17173},
  \href{http://arxiv.org/abs/2305.17173}{{\normalfont\ttfamily
  arXiv:2305.17173}}\relax
\mciteBstWouldAddEndPuncttrue
\mciteSetBstMidEndSepPunct{\mcitedefaultmidpunct}
{\mcitedefaultendpunct}{\mcitedefaultseppunct}\relax
\EndOfBibitem
\bibitem{Amendola:2012ky}
L.~Amendola, M.~Kunz, M.~Motta, I.~D. Saltas, and I.~Sawicki,
  \ifthenelse{\boolean{articletitles}}{\emph{{Observables and unobservables in
  dark energy cosmologies}},
  }{}\href{https://doi.org/10.1103/PhysRevD.87.023501}{\prd \textbf{87} (2013)
  023501}, \href{http://arxiv.org/abs/1210.0439}{{\normalfont\ttfamily
  arXiv:1210.0439}}\relax
\mciteBstWouldAddEndPuncttrue
\mciteSetBstMidEndSepPunct{\mcitedefaultmidpunct}
{\mcitedefaultendpunct}{\mcitedefaultseppunct}\relax
\EndOfBibitem
\bibitem{Amendola:2013qna}
L.~Amendola, S.~Fogli, A.~Guarnizo, M.~Kunz, and A.~Vollmer,
  \ifthenelse{\boolean{articletitles}}{\emph{{Model-independent constraints on
  the cosmological anisotropic stress}},
  }{}\href{https://doi.org/10.1103/PhysRevD.89.063538}{\prd \textbf{89} (2014)
  063538}, \href{http://arxiv.org/abs/1311.4765}{{\normalfont\ttfamily
  arXiv:1311.4765}}\relax
\mciteBstWouldAddEndPuncttrue
\mciteSetBstMidEndSepPunct{\mcitedefaultmidpunct}
{\mcitedefaultendpunct}{\mcitedefaultseppunct}\relax
\EndOfBibitem
\bibitem{Euclid:2019clj}
Euclid, A.~Blanchard, S.~Camera, C.~Carbone, V.~F. Cardone, S.~Casas {\em
  et~al.}, \ifthenelse{\boolean{articletitles}}{\emph{{Euclid preparation: VII.
  Forecast validation for Euclid cosmological probes}},
  }{}\href{https://doi.org/10.1051/0004-6361/202038071}{\aap \textbf{642}
  (2020) A191}, \href{http://arxiv.org/abs/1910.09273}{{\normalfont\ttfamily
  arXiv:1910.09273}}\relax
\mciteBstWouldAddEndPuncttrue
\mciteSetBstMidEndSepPunct{\mcitedefaultmidpunct}
{\mcitedefaultendpunct}{\mcitedefaultseppunct}\relax
\EndOfBibitem
\bibitem{Gleyzes:2015rua}
J.~Gleyzes, D.~Langlois, M.~Mancarella, and F.~Vernizzi,
  \ifthenelse{\boolean{articletitles}}{\emph{{Effective Theory of Dark Energy
  at Redshift Survey Scales}},
  }{}\href{https://doi.org/10.1088/1475-7516/2016/02/056}{\jcap \textbf{02}
  (2016) 056}, \href{http://arxiv.org/abs/1509.02191}{{\normalfont\ttfamily
  arXiv:1509.02191}}\relax
\mciteBstWouldAddEndPuncttrue
\mciteSetBstMidEndSepPunct{\mcitedefaultmidpunct}
{\mcitedefaultendpunct}{\mcitedefaultseppunct}\relax
\EndOfBibitem
\bibitem{Raveri:2021dbu}
M.~Raveri, L.~Pogosian, K.~Koyama, M.~Martinelli, A.~Silvestri {\em et~al.},
  \ifthenelse{\boolean{articletitles}}{\emph{{Principal reconstructed modes of
  dark energy and gravity}},
  }{}\href{https://doi.org/10.1088/1475-7516/2023/02/061}{\jcap \textbf{02}
  (2023) 061}, \href{http://arxiv.org/abs/2107.12990}{{\normalfont\ttfamily
  arXiv:2107.12990}}\relax
\mciteBstWouldAddEndPuncttrue
\mciteSetBstMidEndSepPunct{\mcitedefaultmidpunct}
{\mcitedefaultendpunct}{\mcitedefaultseppunct}\relax
\EndOfBibitem
\bibitem{DES:2021bpo}
DES, A.~Porredon {\em et~al.}, \ifthenelse{\boolean{articletitles}}{\emph{{Dark
  Energy Survey Year 3 results: Cosmological constraints from galaxy clustering
  and galaxy-galaxy lensing using the MagLim lens sample}},
  }{}\href{https://doi.org/10.1103/PhysRevD.106.103530}{Phys.\ Rev.\ D
  \textbf{106} (2022) 103530},
  \href{http://arxiv.org/abs/2105.13546}{{\normalfont\ttfamily
  arXiv:2105.13546}}\relax
\mciteBstWouldAddEndPuncttrue
\mciteSetBstMidEndSepPunct{\mcitedefaultmidpunct}
{\mcitedefaultendpunct}{\mcitedefaultseppunct}\relax
\EndOfBibitem
\bibitem{Fang:2019xat}
X.~Fang, E.~Krause, T.~Eifler, and N.~MacCrann,
  \ifthenelse{\boolean{articletitles}}{\emph{{Beyond Limber: Efficient
  computation of angular power spectra for galaxy clustering and weak
  lensing}}, }{}\href{https://doi.org/10.1088/1475-7516/2020/05/010}{JCAP
  \textbf{05} (2020) 010},
  \href{http://arxiv.org/abs/1911.11947}{{\normalfont\ttfamily
  arXiv:1911.11947}}\relax
\mciteBstWouldAddEndPuncttrue
\mciteSetBstMidEndSepPunct{\mcitedefaultmidpunct}
{\mcitedefaultendpunct}{\mcitedefaultseppunct}\relax
\EndOfBibitem
\bibitem{Camera:2022qfk}
S.~Camera, \ifthenelse{\boolean{articletitles}}{\emph{{A novel method for
  unbiased measurements of growth with cosmic shear}},
  }{}\href{http://arxiv.org/abs/2206.03499}{{\normalfont\ttfamily
  arXiv:2206.03499}}\relax
\mciteBstWouldAddEndPuncttrue
\mciteSetBstMidEndSepPunct{\mcitedefaultmidpunct}
{\mcitedefaultendpunct}{\mcitedefaultseppunct}\relax
\EndOfBibitem
\bibitem{Zuntz:2014csq}
J.~Zuntz, M.~Paterno, E.~Jennings, D.~Rudd, A.~Manzotti {\em et~al.},
  \ifthenelse{\boolean{articletitles}}{\emph{{CosmoSIS: modular cosmological
  parameter estimation}},
  }{}\href{https://doi.org/10.1016/j.ascom.2015.05.005}{Astron.\ Comput.\
  \textbf{12} (2015) 45},
  \href{http://arxiv.org/abs/1409.3409}{{\normalfont\ttfamily
  arXiv:1409.3409}}\relax
\mciteBstWouldAddEndPuncttrue
\mciteSetBstMidEndSepPunct{\mcitedefaultmidpunct}
{\mcitedefaultendpunct}{\mcitedefaultseppunct}\relax
\EndOfBibitem
\bibitem{DES:2020ajx}
DES, A.~Porredon {\em et~al.}, \ifthenelse{\boolean{articletitles}}{\emph{{Dark
  Energy Survey Year 3 results: Optimizing the lens sample in a combined galaxy
  clustering and galaxy-galaxy lensing analysis}},
  }{}\href{https://doi.org/10.1103/PhysRevD.103.043503}{Phys.\ Rev.\ D
  \textbf{103} (2021) 043503},
  \href{http://arxiv.org/abs/2011.03411}{{\normalfont\ttfamily
  arXiv:2011.03411}}\relax
\mciteBstWouldAddEndPuncttrue
\mciteSetBstMidEndSepPunct{\mcitedefaultmidpunct}
{\mcitedefaultendpunct}{\mcitedefaultseppunct}\relax
\EndOfBibitem
\bibitem{Sheldon:2017szh}
E.~S. Sheldon and E.~M. Huff,
  \ifthenelse{\boolean{articletitles}}{\emph{{Practical Weak Lensing Shear
  Measurement with Metacalibration}},
  }{}\href{https://doi.org/10.3847/1538-4357/aa704b}{Astrophys.\ J.\
  \textbf{841} (2017) 24},
  \href{http://arxiv.org/abs/1702.02601}{{\normalfont\ttfamily
  arXiv:1702.02601}}\relax
\mciteBstWouldAddEndPuncttrue
\mciteSetBstMidEndSepPunct{\mcitedefaultmidpunct}
{\mcitedefaultendpunct}{\mcitedefaultseppunct}\relax
\EndOfBibitem
\bibitem{Blazek:2017wbz}
J.~Blazek, N.~MacCrann, M.~A. Troxel, and X.~Fang,
  \ifthenelse{\boolean{articletitles}}{\emph{{Beyond linear galaxy
  alignments}}, }{}\href{https://doi.org/10.1103/PhysRevD.100.103506}{Phys.\
  Rev.\ D \textbf{100} (2019) 103506},
  \href{http://arxiv.org/abs/1708.09247}{{\normalfont\ttfamily
  arXiv:1708.09247}}\relax
\mciteBstWouldAddEndPuncttrue
\mciteSetBstMidEndSepPunct{\mcitedefaultmidpunct}
{\mcitedefaultendpunct}{\mcitedefaultseppunct}\relax
\EndOfBibitem
\bibitem{Bridle:2007ft}
S.~Bridle and L.~King, \ifthenelse{\boolean{articletitles}}{\emph{{Dark energy
  constraints from cosmic shear power spectra: impact of intrinsic alignments
  on photometric redshift requirements}},
  }{}\href{https://doi.org/10.1088/1367-2630/9/12/444}{New J.\ Phys.\
  \textbf{9} (2007) 444},
  \href{http://arxiv.org/abs/0705.0166}{{\normalfont\ttfamily
  arXiv:0705.0166}}\relax
\mciteBstWouldAddEndPuncttrue
\mciteSetBstMidEndSepPunct{\mcitedefaultmidpunct}
{\mcitedefaultendpunct}{\mcitedefaultseppunct}\relax
\EndOfBibitem
\bibitem{Handley:2015fda}
W.~J. Handley, M.~P. Hobson, and A.~N. Lasenby,
  \ifthenelse{\boolean{articletitles}}{\emph{{PolyChord: nested sampling for
  cosmology}}, }{}\href{https://doi.org/10.1093/mnrasl/slv047}{Mon.\ Not.\
  Roy.\ Astron.\ Soc.\  \textbf{450} (2015) L61},
  \href{http://arxiv.org/abs/1502.01856}{{\normalfont\ttfamily
  arXiv:1502.01856}}\relax
\mciteBstWouldAddEndPuncttrue
\mciteSetBstMidEndSepPunct{\mcitedefaultmidpunct}
{\mcitedefaultendpunct}{\mcitedefaultseppunct}\relax
\EndOfBibitem
\bibitem{Lewis:2019xzd}
A.~Lewis, \ifthenelse{\boolean{articletitles}}{\emph{{GetDist: a Python package
  for analysing Monte Carlo samples}},
  }{}\href{http://arxiv.org/abs/1910.13970}{{\normalfont\ttfamily
  arXiv:1910.13970}}\relax
\mciteBstWouldAddEndPuncttrue
\mciteSetBstMidEndSepPunct{\mcitedefaultmidpunct}
{\mcitedefaultendpunct}{\mcitedefaultseppunct}\relax
\EndOfBibitem
\bibitem{Lewis:1999bs}
A.~Lewis, A.~Challinor, and A.~Lasenby,
  \ifthenelse{\boolean{articletitles}}{\emph{{Efficient computation of CMB
  anisotropies in closed FRW models}},
  }{}\href{https://doi.org/10.1086/309179}{Astrophys.\ J.\  \textbf{538} (2000)
  473}, \href{http://arxiv.org/abs/astro-ph/9911177}{{\normalfont\ttfamily
  arXiv:astro-ph/9911177}}\relax
\mciteBstWouldAddEndPuncttrue
\mciteSetBstMidEndSepPunct{\mcitedefaultmidpunct}
{\mcitedefaultendpunct}{\mcitedefaultseppunct}\relax
\EndOfBibitem
\bibitem{Bellini:2014fua}
E.~Bellini and I.~Sawicki, \ifthenelse{\boolean{articletitles}}{\emph{{Maximal
  freedom at minimum cost: linear large-scale structure in general
  modifications of gravity}},
  }{}\href{https://doi.org/10.1088/1475-7516/2014/07/050}{\jcap \textbf{07}
  (2014) 050}, \href{http://arxiv.org/abs/1404.3713}{{\normalfont\ttfamily
  arXiv:1404.3713}}\relax
\mciteBstWouldAddEndPuncttrue
\mciteSetBstMidEndSepPunct{\mcitedefaultmidpunct}
{\mcitedefaultendpunct}{\mcitedefaultseppunct}\relax
\EndOfBibitem
\bibitem{Gleyzes:2015pma}
J.~Gleyzes, D.~Langlois, M.~Mancarella, and F.~Vernizzi,
  \ifthenelse{\boolean{articletitles}}{\emph{{Effective Theory of Interacting
  Dark Energy}}, }{}\href{https://doi.org/10.1088/1475-7516/2015/08/054}{\jcap
  \textbf{08} (2015) 054},
  \href{http://arxiv.org/abs/1504.05481}{{\normalfont\ttfamily
  arXiv:1504.05481}}\relax
\mciteBstWouldAddEndPuncttrue
\mciteSetBstMidEndSepPunct{\mcitedefaultmidpunct}
{\mcitedefaultendpunct}{\mcitedefaultseppunct}\relax
\EndOfBibitem
\bibitem{code}
I.~Tutusaus, C.~Bonvin, and N.~Grimm,
  \ifthenelse{\boolean{articletitles}}{\emph{{Measurement of the Weyl potential
  evolution from the first three years of Dark Energy Survey data}}, }{} 2024.
\newblock
  doi:~\href{https://doi.org/10.5281/zenodo.13734927}{10.5281/zenodo.13734927}\relax
\mciteBstWouldAddEndPuncttrue
\mciteSetBstMidEndSepPunct{\mcitedefaultmidpunct}
{\mcitedefaultendpunct}{\mcitedefaultseppunct}\relax
\EndOfBibitem
\bibitem{DES:2021zxv}
DES, S.~Pandey {\em et~al.}, \ifthenelse{\boolean{articletitles}}{\emph{{Dark
  Energy Survey year 3 results: Constraints on cosmological parameters and
  galaxy-bias models from galaxy clustering and galaxy-galaxy lensing using the
  redMaGiC sample}},
  }{}\href{https://doi.org/10.1103/PhysRevD.106.043520}{Phys.\ Rev.\ D
  \textbf{106} (2022) 043520},
  \href{http://arxiv.org/abs/2105.13545}{{\normalfont\ttfamily
  arXiv:2105.13545}}\relax
\mciteBstWouldAddEndPuncttrue
\mciteSetBstMidEndSepPunct{\mcitedefaultmidpunct}
{\mcitedefaultendpunct}{\mcitedefaultseppunct}\relax
\EndOfBibitem
\bibitem{DES:2020sjz}
DES, R.~Cawthon {\em et~al.}, \ifthenelse{\boolean{articletitles}}{\emph{{Dark
  Energy Survey Year 3 results: calibration of lens sample redshift
  distributions using clustering redshifts with BOSS/eBOSS}},
  }{}\href{https://doi.org/10.1093/mnras/stac1160}{Mon.\ Not.\ Roy.\ Astron.\
  Soc.\  \textbf{513} (2022) 5517},
  \href{http://arxiv.org/abs/2012.12826}{{\normalfont\ttfamily
  arXiv:2012.12826}}\relax
\mciteBstWouldAddEndPuncttrue
\mciteSetBstMidEndSepPunct{\mcitedefaultmidpunct}
{\mcitedefaultendpunct}{\mcitedefaultseppunct}\relax
\EndOfBibitem
\end{mcitethebibliography}

\end{document}